\documentclass[
 reprint,
superscriptaddress,
 amsmath,amssymb,
 aps,
]{revtex4-2}

\usepackage{graphicx}
\usepackage{dcolumn}
\usepackage{bm}
\usepackage{hyperref}

\hypersetup{
    colorlinks,
    linkcolor={blue!90!black},
    citecolor={blue!90!black},
    urlcolor={blue!90!black}
}

\usepackage[all]{hypcap}

\usepackage{enumitem}

\usepackage{graphicx}
\usepackage{xcolor}
\usepackage{mathrsfs}
\usepackage{dsfont}
\usepackage{accents}
\usepackage{textcomp}
\usepackage{float}

\begin{document}

\preprint{APS/123-QED}

\title{  L\'evy distributed fluctuations in the living cell cortex  }

\author{Shankar Sivarajan}
\affiliation{
 Department of Physics and Astronomy, Johns Hopkins University, Baltimore, Maryland }

\author{Yu Shi}
\affiliation{
 Department of Physics and Astronomy, Johns Hopkins University, Baltimore, Maryland }

\author{Katherine M. Xiang}
\affiliation{
 Department of Physics and Astronomy, Johns Hopkins University, Baltimore, Maryland }

\author{Clary Rodr{\'i}guez-Cruz}
\affiliation{
 Department of Chemical and Biomolecular Engineering, University of Pennsylvania, Philadelphia, Pennsylvania }

\author{Christopher L. Porter}%
 \affiliation{
 Department of Chemical and Biomolecular Engineering, University of Pennsylvania, Philadelphia, Pennsylvania }

\author{Geran M. Kostecki}%
 \affiliation{
 Department of Biomedical Engineering, Johns Hopkins University, Baltimore, Maryland}

 \author{Leslie Tung}%
 \affiliation{
 Department of Biomedical Engineering, Johns Hopkins University, Baltimore, Maryland}

\author{John C. Crocker}%
 \email{jcrocker@seas.upenn.edu}
 \affiliation{
 Department of Chemical and Biomolecular Engineering, University of Pennsylvania, Philadelphia, Pennsylvania }

\author{Daniel H. Reich}%
 \email{reich@jhu.edu}
 \affiliation{
 Department of Physics and Astronomy, Johns Hopkins University, Baltimore, Maryland }

\date{\today}

\begin{abstract}
The actomyosin cortex is an active material that provides animal cells with a strong but flexible exterior, whose mechanics, including non-Gaussian fluctuations and occasional large displacements or cytoquakes, have defied explanation.  We study the active fluctuations of the cortex using nanoscale tracking of  arrays of flexible microposts adhered to multiple cultured cell types.  When the confounding effects of static heterogeneity and tracking error are removed, the fluctuations are found to be heavy-tailed and well-described by a truncated L\'evy alpha-stable distribution over a wide range of timescales, in multiple cell types.  The largest random displacements closely resemble the earlier-reported cytoquakes, but notably, we find these cytoquakes are not due to  earthquake-like cooperative rearrangement of many cytoskeletal elements.  Rather, they are indistinguishable from chance large excursions of a super-diffusive random process driven by heavy-tailed noise.  The non-cooperative microscopic events driving these fluctuations need not be larger than the expected elastic energy of single tensed cortical actin filaments, and the implied distribution of microscopic event energies will need to be accounted for by future models of the cytoskeleton.
\end{abstract}

\keywords{ Cytoskeleton $|$ Cell Mechanics $|$ Cortex $|$ Active Matter $|$ Cytoquakes} 



\maketitle

\section{\label{sec:Intro} Introduction}

The actomyosin cortex is a thin sheet of active matter formed from actin filaments, crosslinking proteins and myosin contractile motors existing in a dynamic steady state.  After decades of study both in cells \cite{fabry2001scaling, hoffman2006consensus, solon_fibroblast_2007, hoffman2009cell, tee_cell_2011, vargas-pinto_effect_2013, rigato_high-frequency_2017, shi2019dissecting} and in reconstituted gels \cite{gardel2004elastic, mizuno_nonequilibrium_2007}, the mechanical properties of the actomyosin cortex are well known; it is a tensed, nearly elastic network that resists deformations via a dynamic shear modulus which is a weak power-law of frequency \cite{fabry2001scaling, hoffman2006consensus, trepat2007universal, hoffman2009cell, balland2006power, mandadapu2008cytoskeleton, rigato_high-frequency_2017, shi2019dissecting}. This sheet undergoes active fluctuations that are super-diffusive \cite{caspi2000enhanced, lau2003microrheology, guo2014probing} and heavy-tailed \cite{bursac2005cytoskeletal,alencar2016non, shi2019dissecting}; but such measurements are often confounded by heterogeneity effects.  These varied phenomena have not yet been reproduced by a physics-based model.  Some models \cite{broedersz2010cross, mulla2019origin, chen2021nonlinear} predict power-law shear moduli, but do not explain the non-Gaussian fluctuations.   While Fredberg and colleagues \cite{fabry2001scaling,bursac2005cytoskeletal} have long noted similarities between cytoskeletal networks and soft glassy materials (SGMs) such as foams and emulsions \cite{hebraud1998mode,sollich1997rheology,sollich1998rheological, hwang2016understanding}, with both displaying power-law rheology \cite{hebraud1998mode,sollich1997rheology,sollich1998rheological, hwang2016understanding, giavazzi2020multiple,lavergne2022delayed, rodriguez2022fractal}, super-diffusive dynamics \cite{hwang2016understanding, giavazzi2020multiple,lavergne2022delayed, rodriguez2022fractal} and non-Gaussian displacements \cite{rodriguez2022fractal}, how foams and the cytoskeleton could obey the same physics remains an open question.  More recently, the cortex has also been observed to undergo occasional large displacements in the plane \cite{alencar2016non,shi2019dissecting, shi2021pervasive}, termed cytoquakes, which due to their power-law displacement distribution and large energy scales are hypothesied to be the result of earthquake-like cooperative rearrangement of many cytoskelatal elements.  Such cooperative motion suggests the cortex may exist near a mechanical critical point, on the cusp of instability. Testing of different cytoquake models \cite{liman2020role, floyd2021understanding, swartz2021active}, however, is currently limited by the available data. 

Here we report low noise, high statistical power measurements of the lateral fluctuations and cytoquakes of the actomyosin cortex of multiple cell types.  We used micropost array detectors (mPADs) consisting of dozens of flexible microposts anchored to cells' basal cortex. The data from each micropost was rescaled to correct for post-to-post heterogenity, before being pooled together.  While the largest post displacements resemble the previously reported cytoquakes, we find that their functional form is consistent with a stationary random process that is unlike that in earthquakes and other non-stationary (time-dependent) random processes.   We find that the entire distribution of micropost displacements, from the nanoscale to the cytoquake scale, is well described by a L\'evy alpha-stable distribution with an exponential truncation, resembling a recently described, stationary super-diffusive random process termed Linear Fractional Stable Motion (LFSM) \cite{stoev2004simulation, burnecki2010fractional}.  Physically, such a process corresponds to a viscoelastic solid driven by non-Gaussian noise having a heavy-tailed amplitude distribution, and naturally explains the previously reported power-law distribution of cortical displacements.  This finding also provides a potential explanation for the similarity of the cortex and SGMs \cite{giavazzi2020multiple, lavergne2022delayed}---both may  be viscoelastic solids driven by heavy-tailed active noise \cite{rodriguez2022fractal}.  Last, we estimate that the maximum energy of the microscopic processes driving the cortical fluctuations may be smaller than the energy in single tensed cortical actin filaments.  Known microscopic energy release processes occurring during the rapid turnover of the cortex \cite{fritzsche2013analysis}  appear sufficient to explain our fluctuation measurements and cytoquake observations without earthquake-like cooperativity. Our measurements are likely to be sensitive to cortical details such as the cortical actin filament energy distribution, related to filament tension and length, and may be useful to constrain future physical models of the actomysin cortex.

\begin{figure}
		\centering
        \includegraphics[width = \columnwidth]{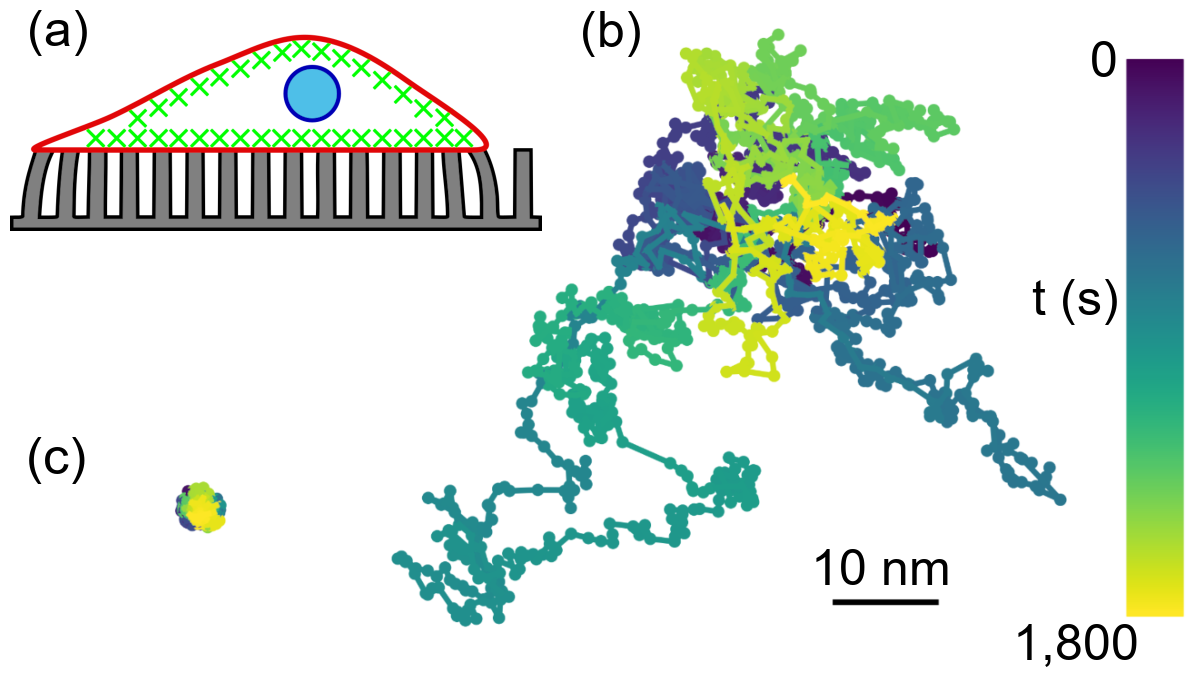}
		\caption{ (a) Schematic of a cell on a micropost array. Cellular traction forces deflect the posts, which are tracked optically.   (b) The trajectory of a micropost ($ k = 15.8$ nN/$\mu$m) coupled to the cortex of a 3T3 cell, over 1,800 s, after time-averaging data recorded at 10 fps to 1 fps.  (c) The trajectory of a micropost not coupled to the cell shows the measurement error.}
		\label{fig:illustration}
\end{figure}

\section{\label{sec:Intro} Experimental Methods}

mPAD arrays are a well-established technique for quantifying cell mechanics and contractility \cite{tan2003cells, du2005force, sniadecki2007magnetic, geng2016review, wolfenson2016tropomyosin, fu2010mechanical}. A schematic of a cell on an mPAD array is shown in Fig.~\ref{fig:illustration}(a). Recently, we have demonstrated \cite{shi2019dissecting, shi2021pervasive, shi2022methods} that mPADs can be used to quantify cortical fluctuations with sub-nm precision. Building on this approach, for this study we used a poly(dimethylsiloxane) (PDMS) mPAD device platform \cite{shi2019dissecting, shi2021pervasive, shi2022methods} consisting of 1.8 $\mu$m diameter microposts on hexagonal lattices with center-to-center spacing 4 $\mu$m.
Micropost heights of 9.1, 6.4, and  5.7 $\mu$m were used, providing effective spring constants for small lateral deflections $k$ of 5.5, 15.8 and 22.3 nN/$\mu$m \cite{fu2010mechanical},
corresponding to substrate stiffnesses 4.3, 12, and 17 kPa \cite{weng2011synergistic}. (We will refer to these as `low' (L), `medium' (M) and `high' (H) stiffness substrates.) The mPAD devices were functionalized to restrict cell adhesion to the micropost tops \cite{tan2003cells} (See SI Methods \cite{supplementary}.)

NIH 3T3 fibroblasts (ATCC), human embryonic kidney (HEK) cells (ATCC), human bone osteosarcoma epithelial cells (U2OS, gift from T. Schroer, JHU), and primary neonatal rat cardiac fibroblasts (CFs) were cultured and seeded on the mPADs as described previously \cite{shi2021pervasive}. Cardiac myofibroblasts (MFs) were produced from the CFs by treatment with TGF-$\beta$1 \cite{shi2021pervasive} prior to seeding.  Bright field videos of individual cells, of duration 30 minutes, were recorded at 10 frames per second (fps) or 100 fps (HEK and U2OS cells only)  \cite{shi2019dissecting, shi2022methods} (See SI Methods \cite{supplementary}.).  A previous study \cite{shi2019dissecting} suggests that the cortex is stiffer than the micropost spring, allowing us to interpret post deflections as cortical displacements rather than forces. While the posts' spring restoring force on the cortical displacements must limit the displacement amplitude, such effects have been shown to be small on the time scales studied here \cite{shi2019dissecting}.  

The trajectories of the posts $\mathbf{r}(t)$ were determined by a centroid tracking algorithm \cite{crocker1996methods, shi2019dissecting,shi2022methods}. To improve resolution, the post centroids were time averaged to 1 fps, yielding a positional uncertainty of $\Delta x \approx 0.5$ nm. 
To isolate posts coupled to the cortex, we adapted and refined an approach reported previously \cite{shi2019dissecting, shi2021pervasive, shi2022methods}, using the post trajectories' mean squared displacements (MSDs), average traction force,  displacement range, and non-Gaussian parameter to identify posts coupled to the cell, to distinguish cortically-associated posts from stress-fiber-associated posts, and to screen out data perturbed by out-of-focus debris  (See  SI Methods and Fig. S1  \cite{supplementary}.) 

\begin{figure}
        \includegraphics[width = \columnwidth]{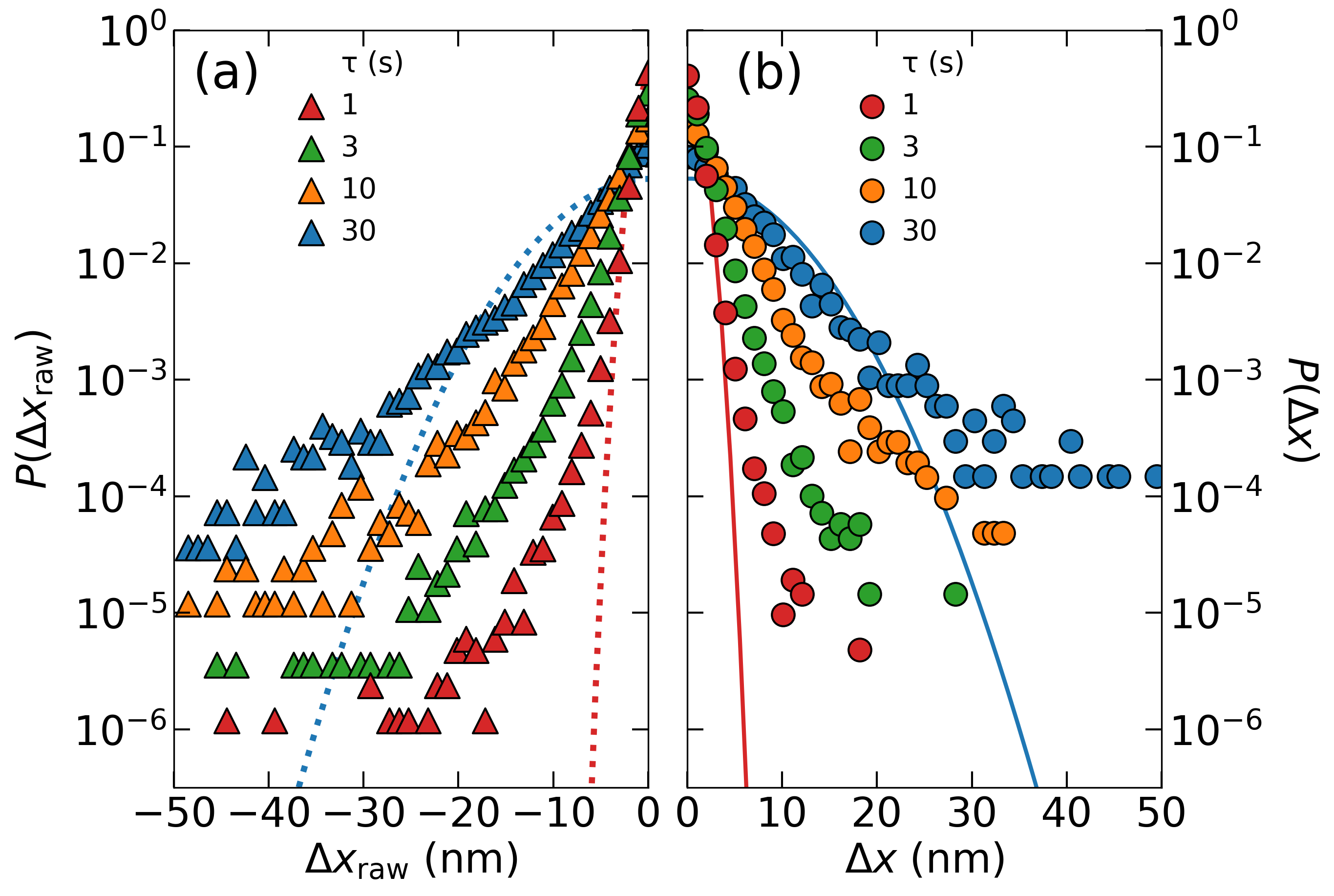}
        
		\caption{The van Hove displacement distribution is heavy-tailed even after post-to-post heterogeneity is removed. (a) Distribution of measured displacements $P(\Delta x_{\textrm{raw}})$ for the $x$-component of the post trajectories at fixed lag time $\tau$  for an ensemble of 336 posts from 10 3T3 cells.  Post stiffness:   $ k = 15.8$ nN/$\mu$m (Medium stiffness). Only the $\Delta x_{\textrm{raw}}\le 0$ half of the symmetric distribution is shown, to facilitate comparison with Panel (b).  (b) Van~Hove distributions $P(\Delta x)$ following rescaling to remove static heterogeneity as described in the text. The data in (b) only include steps from posts in the uppermost quartile of the distribution of the trajectories' geometric means. Only the $\Delta x \ge 0$ half of the symmetric distribution is shown. The lines in (a) and (b) are best-fit Gaussians at $\tau = 1\,\textrm{s}$ and $\tau = 30\,\textrm{s}$.}
		\label{fig:linlog}
\end{figure}

\section{\label{sec:MPADS} MPADs reveal nanometer scale motion of the cortex}

Microposts attached to the cell cortex showed dynamic displacements on the 10 nm scale, driven by lateral cortical fluctuations, while background posts did not (Fig.~\ref{fig:illustration}(b),(c)). To characterize these random post/cortical displacements, we computed the van Hove self correlation function, or the probability distribution of displacements $P(\Delta x_{\textrm{raw}}(\tau))$ in a waiting (or lag) time $\tau$.  Figure \ref{fig:linlog}(a) shows the displacement distribution for an ensemble of 336 posts from 9 different 3T3 cells, for   1~s $\le \tau \le $ 30~s.  At all lag times measured, the distributions were non-Gaussian with pronounced heavy tails at large $|\Delta x_{\textrm{raw}}|$. Such measurements, which pool results over many posts,  are susceptible to confounding effects due to heterogeneity \cite{trepat2007universal}, such as variations in how different posts/tracers are coupled to the cytoskeleton, or differences between cells.  Prior studies of cells' cortical displacements using single tracers \cite{alencar2016non, shi2019dissecting}, have yielded similar results to Fig.~\ref{fig:linlog}(a), suggesting that such fluctuations are not due solely to heterogeneity.  

A pooled measurement of displacements can be corrected for heterogeneity by rescaling. In the case of \textit{static heterogeneity} each post `$i$' reports the time-dependent motion $x^i_{\textrm{raw}}(t)$ of a segment of the cortex multiplied by a different, time-independent constant. This suggests that each post's motion can be rescaled by a single constant so as to have the same typical amplitude as the entire ensemble, \textit{e.g.}

\begin{equation}\label{eq:gmscaling}
\Delta x^i(t) =\Delta x^i_{\textrm{raw}}(t)\left[\dfrac{
\langle \textrm{GM}(|\Delta x^i_{\textrm{raw}}(\tau = 10\,\textrm{s})|) \rangle_{\textrm{ens}}}{\textrm{GM}(|\Delta x^i_{\textrm{raw}}(\tau = 10\,\textrm{s})|)} \right],
\end{equation}

\noindent where $\textrm{GM}( . )$ designates the geometric mean of a set of numbers, a robust measure of the typical value in heavy-tailed distributions, and $\langle . \rangle_{\textrm{ens}}$ is an ensemble average over all posts.  The choice to scale by the displacements  at $\tau = 10$ s minimized the effects of measurement error while retaining good statistics.  The distributions of post rescaling factors for our set of cell types and substrate stiffnesses are shown in  Fig.~S2  \cite{supplementary}.  A control analysis showed that the rescaling factor was not significantly time dependent over our $30$ minute datasets; cell to cell variations were also small \cite{supplementary}.

A potential concern with this rescaling procedure is the amplification of noise contributions from posts with small GMs. In addition to non-biological sources of noise, such as camera noise, which can be quantified using the background posts, a second noise source can arise from fluctuations in the local optical density over a post due to internal cellular rearrangements \cite{shi2019dissecting}. From an assessment of the magnitude of such contributions (See SI and Fig.~S3  \cite{supplementary}), we determined that these effects do not contribute significantly to the signal from posts whose GM is in the top quartile of the GM distribution, and so we used those posts for subsequent analysis.

Figure \ref{fig:linlog}(b) shows the resulting pooled distribution of cortical displacements for 3T3 cells, corrected for static heterogeneity, for a range of lag times.  This confirms that cortical fluctuations are intrinsically non-Gaussian, while our statistical power allows us to observe that such non-Gaussianity persists to long lag times. At the shortest lag time measured, $\tau = 1$ s, we find random displacements that are up to 40 times larger than the geometric mean, similar to reports for cytoquakes \cite{alencar2016non}. The results for our other experimental configurations and cell types, shown in  Fig.~S4  \cite{supplementary}, show similar but somewhat less non-Gaussian behavior. 

Visual examination of the trajectories corresponding to the largest displacements showed two qualitatively different kinds of events. The majority of the events were isotropically directed and showed roughly sigmoidal shapes with typical widths $>1$ s, while other events were very abrupt, with typical widths $<1$ s, and were strongly directed toward the posts' resting locations. Examples of such trajectories and their respective angular probability distributions are shown in  Fig.~S5(a)-(d)  \cite{supplementary}. Given the tendency of the abrupt events to move toward the post's resting location, we hypothesize that they are due to transient loading of the cortex by stress fibers followed by detachment and/or de-adhesion events of the micropost from the cell \cite{shi2019dissecting}.  Given that such processes are not fluctuations within the cortex itself, as might occur due to motor activity or remodeling, we screened the abrupt events out from our analyses below, except where noted.

\section{\label{sec:IEF} Cytoquakes do not resemble earthquake-like events}

\begin{figure}[!ht]
		\centering
       \includegraphics[width = \columnwidth]{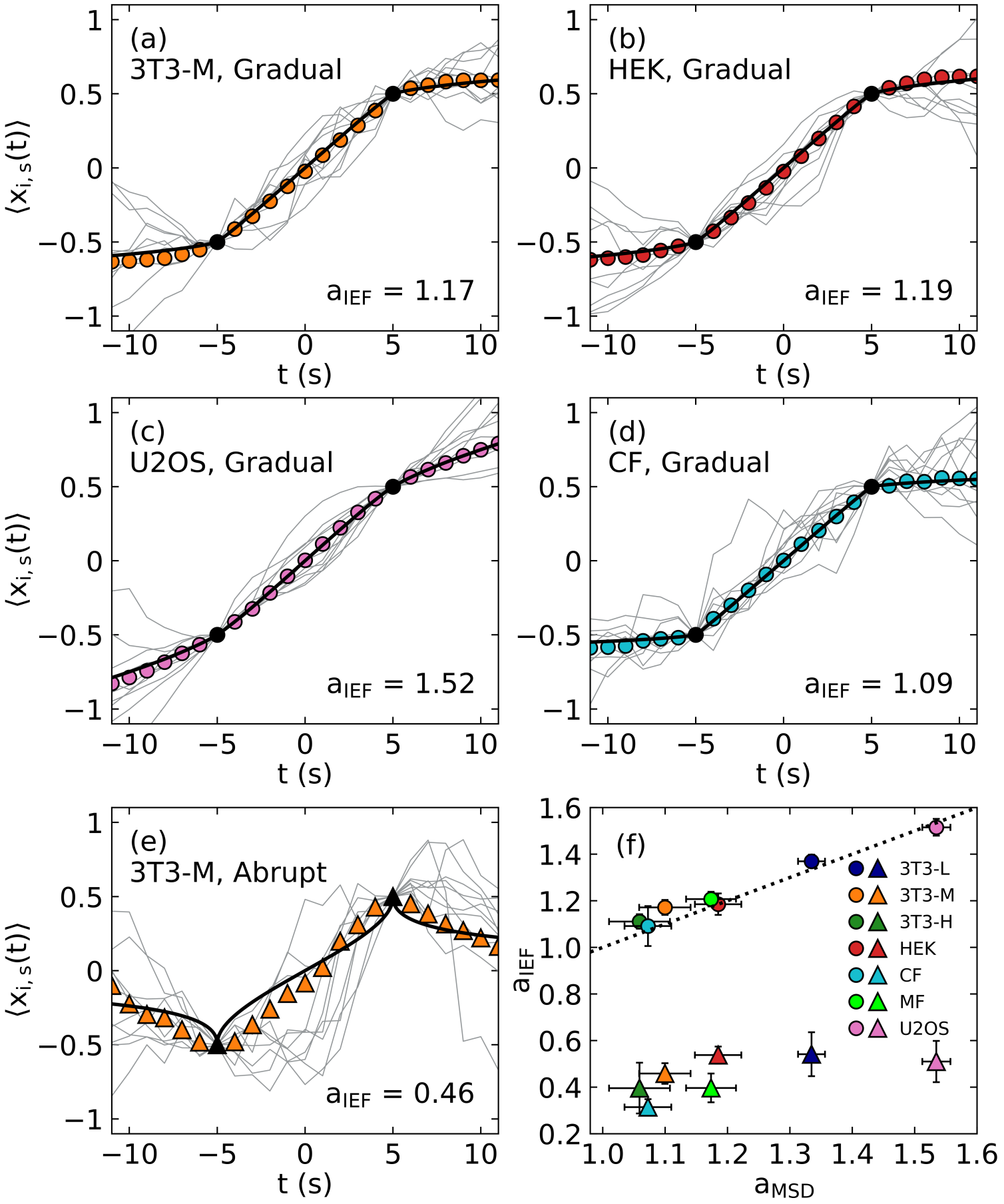}
		\caption{Averaged scaled trajectories $\langle x_{i,s}(t)\rangle$ for large displacements are well-described by the interpolation-extrapolation function (IEF). (a)--(d)~$\langle x_{i,s}(t)\rangle$ for the $200$ largest gradual displacements at $\tau = 10$ s, for 4 different cell types and conditions, as described in the text. Ten individual scaled trajectories $x_{i,s}(t)$ are superimposed in each case (grey).   (e)~$\langle x_{i,s}(t)\rangle$ for the $40$ largest abrupt displacements for the 3T3 cells shown in (a).  Solid lines in (a)--(e) are fits to the IEF, Eq.~\ref{eq:qhoft}. (f)~The exponents $a_{\textrm{IEF}}$ from fitting the gradual IEFs (circles) and the corresponding MSD exponents $a_{\textrm{MSD}}$ show close agreement. Additional datasets and fits to the IEF from  Fig.~S6  \cite{supplementary} are included in (f). The micropost stiffnesses were: 3T3-L and U2OS: $k=$ 5.5 nN/$\mu$m;  3T3-M and HEK: $k=$ 15.8 nN/$\mu$m; 3T3-H, CF and MF: $k=$ 22.3 nN/$\mu$m. Dashed line: $a_{\textrm{IEF}} = a_{\textrm{MSD}}$. The $a_{\textrm{IEF}}$ exponents for the abrupt displacements (triangles) differ widely from $a_{\textrm{MSD}}$.  Error bars are standard errors, estimated from cell-to-cell variations  \cite{supplementary}. The individual trajectories in (a)--(e) were chosen at random from the 50 largest (Panels (a)--(d)) or 20 largest displacements (Panel (e)).}

		\label{fig:ief}
\end{figure}

The large displacement events we observe (filtering out the abrupt events) have both amplitudes and sigmoid-like time-dependence resembling the cytoquakes we and others have observed previously \cite{alencar2016non,shi2019dissecting, shi2021pervasive}.  We seek to apply statistical tests to the cytoquakes' time-dependent data to determine if they resemble cooperative processes like earthquakes and avalanches, perhaps superimposed on a background noise of other mechanical fluctuations.  We begin by averaging short trajectory segments of many large cytoquake displacements together (selected from the heavy tails of the van Hove distribution).  The results $\langle x_{i,s}(t)\rangle$ of scaling and averaging  the 200 largest gradual displacements at $ \tau = 10$ s for four representative cell types are shown in Fig.~\ref{fig:ief}(a)--(d). During this averaging, the trajectories $ x_{i}(t)$ containing each large displacement were rescaled to pass through the points $(-\tau/2,-1/2)$ and $(+\tau/2,+1/2)$. (See SI for calculation details  \cite{supplementary}.) Figure \ref{fig:ief}(e) shows a similar average of 40 abrupt displacements for comparison. The corresponding data for both gradual and abrupt events for these and our other cell types and substrate conditions are shown in  Fig.~S6  \cite{supplementary}. 

For a stationary (non-cooperative) random process, the average of short trajectory segments rescaled as in Fig.~\ref{fig:ief} is predicted by the interpolation-extrapolation function (IEF) \cite{mandelbrot1968fractional}, which gives the conditional expectation value for a random walk constrained to pass through two points (here $(-\tau/2,-1/2)$ and $(+\tau/2,+1/2)$).  If the MSD of the stationary random process  varies as $\textrm{MSD} \sim \tau^a$, then the IEF is given by:
\begin{equation} \label{eq:qhoft}
\textrm{IEF}(t) = \frac{1}{2 \tau^a }\left( \left\lvert t+\frac{\tau}{2}\right\rvert^a  - \left\lvert t - \frac{\tau}{2}\right\rvert^a \right)
\end{equation}
(See SI for a derivation of this expression \cite{supplementary}.) 

The cytoquake events are well fit by the IEF form, Eq.~\ref{eq:qhoft}, with the exponent as a free parameter, Fig.~\ref{fig:ief}(a)--(d), while the abrupt events, Fig.~\ref{fig:ief}(e) do not. This agreement demonstrates that the cytoquakes' apparent time dependence is indistinguishable from what is obtained by merely sampling and averaging the largest excursions of a stationary super-diffusive random process.  Were cytoquakes a non-stationary cooperative process like an earthquake or avalanche, their time dependence, like the abrupt events, would have no reason to fit to the IEF model. The agreement of the IEF model across multiple cell types leads us to conclude that cytoquakes, at least in our data, are simply statistical mirages.  Data for cytoquake and abrupt events in other cell types and conditions reinforce the above conclusions, see Fig.~S6, \cite{supplementary}.

The exponents for different cell types $\{a_\textrm{IEF}\}$ for the gradual events agree well with the corresponding exponents $\{a_\textrm{MSD}\}$ for the MSDs of the full trajectory datasets, Fig.~\ref{fig:ief}(f). (See  Fig.~S7 for MSD fits \cite{supplementary}.)  The agreement of these exponents confirms that the cytoquakes are driven by the same microscopic fluctuations and mechanics driving all cortical fluctuations. In contrast, the exponents for abrupt events differ significantly from $\{a_\textrm{MSD}\}$, Fig.~\ref{fig:ief}(f), confirming that they are due to a cellular process distinct from the other cortical fluctuations, e.g. micropost recoil after detachment from the cortex. 

\begin{figure}
		\centering
        \includegraphics[width = \columnwidth]{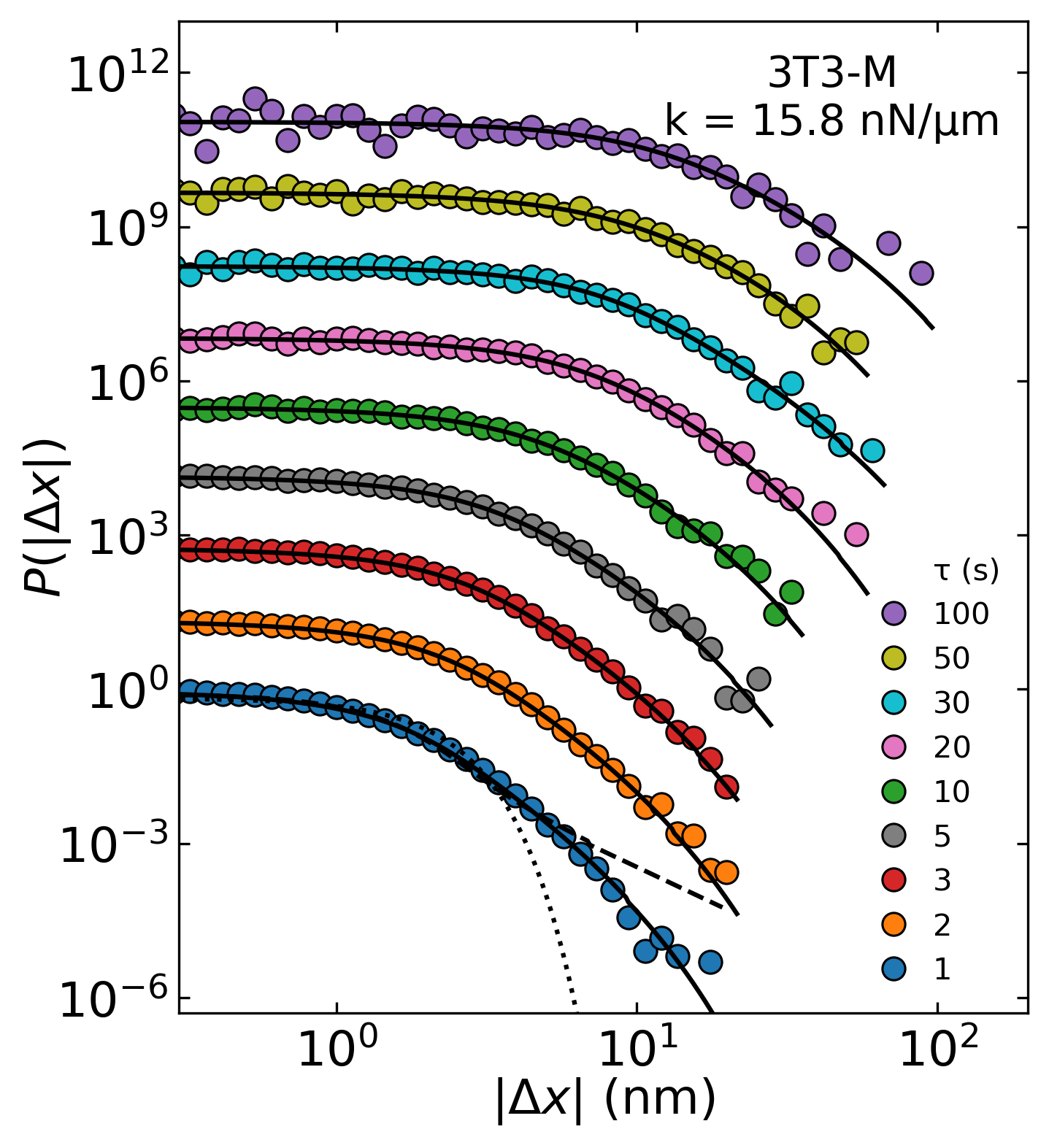}
		\caption{The Van~Hove fluctuation distributions for 3T3 cells are well-described by an ETSD model (solid lines) over lag times $\tau$ from 1--100 s. The best-fit stable distribution (dashed line) shown at $\tau = 1$ s does not capture the behavior near the tail. The best-fit Gaussian is  shown at $\tau = 1$ s for reference (dotted line). The distributions for $\tau > 1$ s are progressively offset by factors of $10^{1.5}$ for clarity.  Micropost stiffness: $ k = 15.8\,\textrm{nm/$\mu$m}$ (Medium stiffness).}
		\label{fig:vanhovesfit}
\end{figure}

\section{\label{sec:ETSD} Displacement distributions have an ETSD form}

Our remaining task is to model the lag-time dependent van Hove distribution that describes the cytoskeletal fluctuations.  To do so, we must first screen out the above described abrupt displacements. As the abrupt displacements were predominantly toward the posts' resting locations, we split the displacements into two sets: one with displacements moving toward the resting location, which we discarded, and the other for those moving away from it, which we retained (Fig.~S5(e)  \cite{supplementary}).  Comparing these two sets of data suggests that the abrupt events (in the `toward' set) are responsible for about 20\% of the total observed fluctuations on a power basis  \cite{supplementary}. 

The cortical displacement distribution with abrupt events removed is shown in Fig.~\ref{fig:vanhovesfit}, plotted in log-log form, for 3T3 cells, and lag times from 1 to 100 s. Other cell types and conditions give similar results, shown in Fig.~S8  \cite{supplementary}.  We find that the cortical fluctuation distribution is well-described by an exponentially-truncated stable distribution (ETSD), given by:
\begin{equation} \label{eq:etsd}
P(\Delta x) = A L_{\alpha}(\Delta x; \alpha,\gamma)e^{-\Delta x/\lambda},
\end{equation}
\noindent where $L_{\alpha}(\Delta x; \alpha,\gamma)$ is the symmetric L\'evy alpha-stable distribution \cite{zaburdaev2015levy} with shape parameter $\alpha$ and scale parameter $\gamma$, $A$ is a normalization constant and $\lambda$ is a truncation length.   The ability to describe the entire distribution with a single function, as opposed to a sum of different functions, confirms our conclusion that we are observing the displacements of a single random process.

\begin{figure}
		\centering
        \includegraphics[width = \columnwidth]{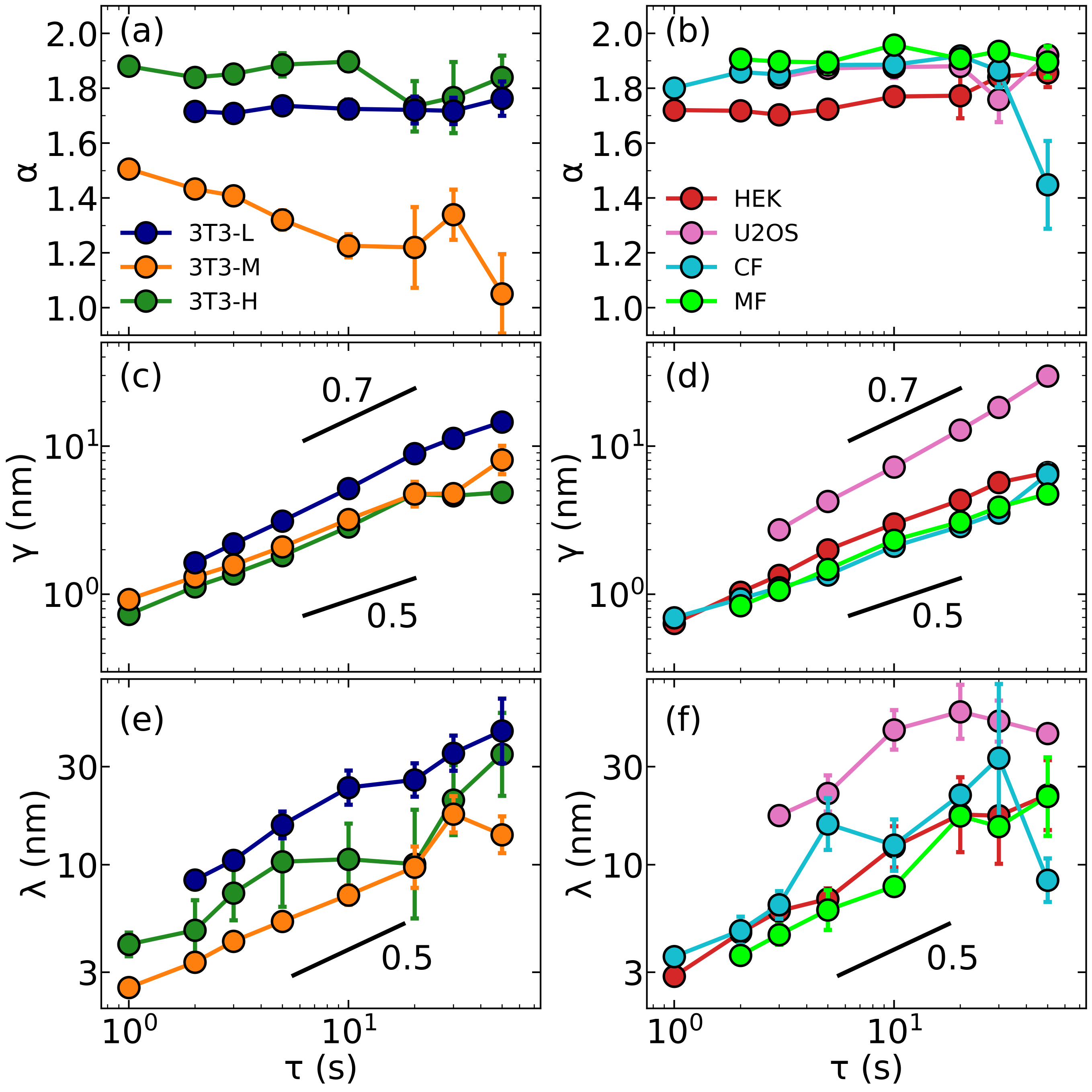}
		\caption{The ETSD fit parameters vs. lag time $\tau$. (a) The shape parameter $\alpha$ varies with substrate stiffness.  (b) The shape parameter $\alpha$ depends on the cell type. (c), (d) The scale parameter $\gamma$ follows a power law with exponent $ \approx 0.7$, corresponding to a super-diffusive MSD. The uncertainty in $\gamma$ is smaller than the size of the markers. (e), (f) The truncation parameter $\lambda$ increases with increasing $\tau$. Error bars are standard deviations of fits to samples drawn from the best-fit  ETSDs to the experimental data.}
		\label{fig:parameters}

\end{figure}

The ETSD parameters for our experiments, summarized in Fig.~\ref{fig:parameters} vary significantly across cell types and post stiffness, with some displaying consistent trends versus lag time.  This suggests that these measures are not `universal', and instead are sensitive to the arrangement of the cortical actin and myosin for different cell types and post stiffnesses. This is in contrast the the universal \emph{rheology} of cells \cite{hoffman2009cell}, and suggests that future actin and myosin perturbation experiments may elucidate the parameters' connections to cortical architecture. 

\section{\label{sec:params} Biophysical interpretation of ETSD parameters}

In this section we report the ETSD parameters and put them into biophysical context. The shape parameter $\alpha(\tau)$, Figs.~\ref{fig:parameters}(a)--(b), describes the exponent of the heavy tail, or equivalently how non-Gaussian the displacements are versus lag time, $\tau$ ($\alpha=2$ is Gaussian).  Six out of seven conditions show  modestly non-Gaussian behavior, $\alpha \approx 1.8$, essentially independent of lag time. Datasets collected on the high stiffness posts ($k = 22$ nN$/\mu m$) show the least non-Gaussian behavior, $\alpha \approx 1.9$. The 3T3 fibroblasts on medium stiffness posts (highlighted in previous figures) show very heavy-tailed displacements having a shape parameter that decreases with lag time.

Heavy-tailed fluctuations in conventional materials are caused by two distinct effects \cite{cipelletti2003universal, stoev2004simulation, burnecki2010fractional}.  The first is geometrical and due to the spatial decay of strain fields around force dipoles \cite{cipelletti2003universal, swartz2021active}.  For the cortical geometry, a thin stiff sheet over a soft interior, this effect contributes negligible non-Gaussianity \cite{swartz2021active}, and predicts $\alpha \approx 2$, inconsistent with our findings.  The second effect is more direct: stable distributed displacements are  the result of microscopic events (e.g. of forces or energy release) that themselves have a heavy-tailed distribution. In other words, non-Gaussian displacements are the result of non-Gaussian `noise' from microscopic processes.  In this case, the tail exponent of the noise amplitude distribution is connected to the stability parameter $\alpha$ of the mesoscopic displacement \cite{stoev2004simulation, burnecki2010fractional} via a generalized Central Limit theorem \cite{uchaikin2011chance}.  Biophysically, our findings suggest that the ETSD distribution of cortical displacements is the result of a heavy-tailed distribution of the random microscopic forces or energy release events driving the fluctuations of the cortex.

The scale parameter $\gamma(\tau)$, of the displacement distributions, Figs.~\ref{fig:parameters}(c)--(d) is akin to the width or `knee' of the distribution and grows with $\tau$, as is evident in Fig.~\ref{fig:vanhovesfit}. The displacement scales are fairly similar across cell types at a given substrate stiffness with, intuitively, larger displacement scales being observed on softer posts. The scale parameters increase as power-laws of lag time, with exponents in the range $0.5- 0.7$.  For a truncated distribution, the scale parameter and standard deviation are correlated, so the mean-squared displacement should scale $\sim \gamma^2$.  Thus our findings appear consistent with the well-known \cite{hoffman2009cell,shi2019dissecting,shi2021pervasive} super-diffusive nature of cortical fluctuations, having non-universal exponents in the range $1.0-1.4$ \cite{hoffman2006consensus, trepat2007universal, guo2014probing, shi2019dissecting,shi2021pervasive}. A classic model of stationary super-diffusive processes is Fractional Brownian Motion (FBM) \cite{mandelbrot1968fractional}, which combines Gaussian noise with a power-law memory kernel.  The FBM model has been extended to include noise having a stable distribution form, termed Linear Fractional Stable Motion (LFSM) \cite{stoev2004simulation, burnecki2010fractional}, a process that displays both super-diffusion and displacements that follow a stable distribution at all lag times.    Notably, an LFSM process creates `power-law' distributed displacements without an earthquake-like mechanism.  Biophysically, a LFSM process would correspond to the cortex being driven by uncorrelated heavy-tailed noise, with a power-law memory kernel provided by the cortex's power-law creep compliance and/or temporal correlations in the noise.
  
The truncation parameter $\lambda(\tau)$, Figs.~\ref{fig:parameters}(e)--(f), describes the effective maximum size of a displacement that occurs in time $\tau$; larger events are exponentially rare. If the displacements are driven by non-Gaussian noise from microscopic events having a heavy-tailed distribution, then the observed truncation simply suggests that these events have a well-defined maximum size. Such truncation is ubiquitous in physical systems, but not present in existing LFSM models.  We find experimentally that the truncation length increases with $\tau$, as roughly $\sim \tau^{1/2}$. We conjecture that such scaling corresponds to a diffusion-like process: $\lambda(\tau) = \lambda_0 \sqrt{k_{max} \tau}$, where $\lambda_0$ is the displacement due to a maximum-sized single microscopic event, and $k_{max}$ is the rate at which such events occur. In this picture, at longer lag times many maximal microscopic events have contributed to a maximal micropost displacement, whose uncorrelated effects add together as in a random walk.

The truncation lengthscale $\lambda(\tau)$ we measure describes the displacement of a mesoscopic micropost, and not any structure in the cortex, such as myosin stepping.  As such it is interesting to convert it to the work done on the micropost by the cortex, via the formula for the elastic energy in a spring $\Delta U_{post} = \frac{1}{2} k \Delta x^2$.  The maximum displacement $\Delta x = \lambda(\tau =1  $ s$) = 3$~nm, and $k=15.8 $~nN/$\mu$m yields an energy of $\Delta U_{post} = 7 \times 10^{-20}$~J. This energy represents a lower bound for the largest energy release events driving cortical fluctuations, assuming that a single such event is responsible for a $\lambda$-sized miropost displacement (See SI for details  \cite{supplementary}.)  If our displacements at $\tau=1$ s actually correspond to multiple such maximal events, the microscopic events' energy could be even lower.  It is useful to compare this energy bound to the characteristic elastic energy in cortical elements, such as highly stretched actin filaments \cite{kojimaDirectMeasurementStiffness1994, liuMechanicsFActinCharacterized2002, matsushitaEvaluationExtensionalTorsional2010}. Following a recent computational model for the cortex \cite{cunha2023building} which suggests that individual actin filaments are stretched to very high tensions up to 400 pN (about 2/3 of their breaking tension \cite{tsudaTorsionalRigiditySingle1996}), we can estimate the maximal elastic energy stored in a single cortical actin filament: $\Delta U_{actin} = 20 \times 10^{-20}$~J. (See SI for details  \cite{supplementary}.) Notably, the fact that the estimated lower bound of the ($\lambda$-scale) energy release event $\Delta U_{post}$ is only about a third the value of $\Delta U_{actin}$, implies that even the largest cortical fluctuations we observe could be readily driven by single microscopic processes in the cortex.  

\section{\label{sec:Discussion} Discussion}

The first paper on cytoquakes \cite{alencar2016non} as well as our subsequent papers \cite{shi2019dissecting,shi2021pervasive} hypothesize that they are due to the collective or cooperative reconfiguration of many microscopic degrees of freedom, with one microscopic event triggering many more in the manner of an avalanche or earthquake.  This hypothesis is motivated by the power-law distributed displacement distributions that we observe in detail here. Moreover, the largest reported cytoquakes, often up to 30 nm in ten seconds, correspond to large work done on the micropost, $\Delta U_{quake} = 7 \times 10^{-18}$ J, far larger than the typical energy in a tensed cortical actin filament  (see  SI  \cite{supplementary}).  Such large cytoquake energies are consistent with the hypothesis of many cytoskeleton elements cooperatively rupturing or reconfiguring.  

In contrast, however, a major finding of this study is that cytoquakes closely resemble chance large excursions of a stationary random process driven by non-Gaussian noise (as in Fig.~\ref{fig:ief}).  The time-dependent shape and power-law distribution of the largest cytoquakes can thus be explained as the chance result of many statistically uncorrelated microscopic events having a heavy-tailed distribution of energy, without any earthquake-like cooperativity.  Moreover, analysis of the truncation parameter $\lambda$ suggest that the largest single microscopic energy release events driving cortical motion are comparable to the energy in single tensed actin filaments. Rather than being a mechanical network near some mechanical critical point, subject to `earthquakes', the cytoskeleton may simply be a mechanical network with little or no cooperativity driven by active stresses due to structural changes of single cortical elements.  

Next, we consider possible physical origins of the random process driving both the observed cortical fluctuations having an ETSD form and the occasional cytoquakes. It is well accepted that the mechanical energy in the cortex comes from mysoin II contraction, and molecular stepping has been seen in cortex-adhered microposts associated with large myosin contractile units \cite{wolfenson2016tropomyosin}. FRAP experiments \cite{fritzsche2013analysis} show rapid cortical turnover with the average tensed actin filament lifetime being only 30 seconds. The removal of filaments is primarily due to cofilin severing followed by depolymerization.  This severing process, or similarly, the unbinding of tensed actin filaments from myosin mini-filaments or crosslinking molecules would abruptly release the stored elastic energy in the tensed actin filament and cause the surrounding network to recoil. This recoil energy would have both the correct magnitude to account for our observed fluctuations, and these events would likely have a broad distribution of energies related to the expected broad distribution of filament tensions \cite{cunha2023building}.  A suitably broad distribution could then lead, potentially, to the heavy-tailed noise implied by our ETSD fluctuations.  Alternatively, the distribution of microscopic elastic energies may not be heavy-tailed, and the heavy-tailed active noise may be due to cooperativity among energy release events on the microscale, as seen in recent simulations \cite{floyd2021understanding}.

Our fluctuation measurements suggest that the actomyosin cortex is driven by highly non-Gaussian noise, due to microscopic processes releasing energies less than the typical elastic energy in single cortical actin filaments.  While these microscopic processes may display some cooperativity, such cooperativity is not required to explain the cytoquake phenomenon.  The observed ETSD distribution of the fluctuations is both a clue to direct  the development of new physical models and a challenge for existing ones to reproduce, for example through a broad distribution of elastic energies in actin filaments or other mechanical elements. The ETSD distribution closely resembles that seen in soft glassy materials \cite{rodriguez2022fractal}, adding to the curious similarity between the cytoskeleton and SGMs \cite{fabry2001scaling}, but this could be explained if both are simply viscoelastic solids driven by highly non-Gaussian active noise.   Unlike cell rheology curves, which are parameterized by a single, nearly universal exponent \cite{hoffman2009cell}, the measured lag time dependent ETSD parameters appear sensitive to cell type-specific details of actomyosin assembly.  Further experiments to link these distribution parameters to actomyosin structure and biochemistry should be useful in enabling the refinement of future models.

\section*{\label{sec:Acknowledgments} Acknowledgments} 

We are grateful to B.~Camley for helpful discussions, C.~S.~Chen for donation of mPAD masters and T.~Schroer for donation of U2OS cells. This work was supported by NSF grants PHY-1915193 and PHY-1915174 and NIH grant HL-127087.

\bibliography{references}

\newpage

\thispagestyle{empty}
\setcounter{page}{1}

\onecolumngrid

\centerline{\bfseries \large Supplementary information for ``Lévy distributed fluctuations in the living cell cortex''}

\vskip 0.15in

\centerline{\large{Shankar Sivarajan, Yu Shi, Katherine M.\ Xiang, Clary Rodríguez-Cruz, Christopher L.\ Porter}}
\vskip 0.05in
\centerline{\large{Geran M.\ Kostecki, Leslie Tung, John C.\ Crocker, Daniel H.\ Reich}}

\vskip 0.25in

\twocolumngrid

\renewcommand{\theequation}{S\arabic{equation}}
\renewcommand{\theHequation}{S\arabic{equation}}
\setcounter{equation}{0}

\section{Expanded Methods}

{\it  Micropost array fabrication:} Micropost substrates were formed from poly(dimethylsiloxane) (PDMS, Sylgard) \cite{shi2019dissecting, shi2021pervasive} via replica molding \cite{tan2003cells} using PDMS negative molds produced from silicon masters \cite{fu2010mechanical}.  The tops of the posts were functionalized with fibronectin (Sigma-Aldrich) via microcontact printing to promote specific cellular adhesion and all remaining surfaces of the arrays were passivated with a 0.2\% w/v Pluronic F-127 (Thermo Fisher Scientific) solution to prevent non-specific adhesion \cite{tan2003cells}.  

{\it Cell culture and data acquisition:} All cell lines were cultured as described previously \cite{shi2019dissecting,shi2021pervasive}.  NIH 3T3 fibroblasts and human embryonic kidney (HEK) cells were obtained from ATCC, and human bone osteosarcoma epithelial cells (U2OS) were a gift from T. Schroer. 
Primary cardiac fibroblasts (CFs) were extracted from the hearts of neonatal (2 day old) Sprague Dawley rats (Harlan, Indianapolis, IN, USA). All animal procedures were performed in compliance with guidelines set by the Johns Hopkins Committee on Animal Care and Use and all federal and state laws and regulations \cite{shi2021pervasive}.  The CFs were maintained in a fibroblastic state by treatment with the TGF-$\beta$ receptor I kinase inhibitor SD-208 (Sigma).  Cardiac myofibroblasts (CMFs) were produced from the CFs by treatment with TGF-$\beta$1 (R+D Systems) for 48 h.

All cells were seeded on mPAD devices and incubated overnight prior to measurements to allow them to adhere to and spread on the micropost arrays.
Bright field videos of 30 min duration were recorded with  a Nikon TE-2000E inverted microscope with a 40$\times$ NA $= 0.6$ extra-long working distance air objective as described in detail previously \cite{shi2019dissecting, shi2022methods}. 

{\it Data reduction and 
	identification of cortically-associated microposts:}
For the high-resolution studies in this work, we needed to improve our previously described data reduction and analysis methods \cite{shi2022methods} to measure the microposts' trajectories, to identify the set of posts coupled to the cortex of each cell, to further reduce experimental noise, and to obtain cleaner sets of cortically-associated microposts. These prior methods were used as a first step, with the addition of multi-frame averaging, to obtain a preliminary set of trajectories and micropost identifications.  Briefly, posts were first provisionally assigned as `background' or `cell-associated' based on visual inspection. Then a centroid-based particle-tracking algorithm \cite{crocker1996methods} was used to obtain the position of each post in each video frame. Frame-to-frame drift was provisionally accounted for by subtracting from each post's trajectory the average displacement of the initial set of background posts in each video frame relative to the initial frame. All post trajectories were then time-averaged to 1 fps to reduce imaging noise.   

The individual posts' mean squared displacements (MSDs) were computed. As the cell-associated posts' MSDs showed power-law behavior, $MSD \propto \tau^a$ for one to two decades in $\tau$, the MSD exponents $a$ for each post were obtained by fitting and subtracting the short-$\tau$ noise floor and averaging the slope of the logarithmic time derivative of the resulting `subtracted MSD' between 5 s $ \le \tau \le $ 10  s. The condition $a >0.5$ was used to identify cell-associated posts, and $ a < 0.1$ was used to identify `background' posts not engaged with the cell. To screen out posts that were not engaged with the cell for the entire 1,800 s measurement interval (due to cell motility, for example), we analyzed the MSDs separately for the first and last third of each video. Posts with MSD exponents $< 0.5$ for either of those intervals were eliminated from further consideration.

The undeflected positions of the cell-associated posts were then determined by interpolation based on the positions of the background posts, and the cell posts were then provisionally classified into cortex-associated and stress-fiber associated posts based on their average traction force, as described previously \cite{shi2019dissecting, shi2021pervasive, shi2022methods}.

While the above procedure was sufficient for our prior studies \cite{shi2019dissecting, shi2021pervasive}, the precise measurement of the cortical fluctuations in this work required several improvements. Importantly, a cleaner set of background posts was needed to measure the frame-to-frame drift more accurately, and to determine better the resting locations of the cell-associated posts, which enter the process of separating the cortical posts from those coupled to other cytoskeletal structures, such as stress fibers.  In our existing procedures,  some posts near the periphery of the cells were misclassified. In addition, there were transient optical disturbances that appeared to be caused by debris in the culture medium floating into the field of view, and which could adversely affect the imaging of both cell-associated and background posts. 

To remove these effects, we classified the posts based on (i) their displacement range $ x_{\textrm{range}}$, defined as the range covered by the $x$-coordinate of a post's trajectory, and (ii) the non-Gaussian parameter for the trajectory
\begin{equation}
	\textrm{NGP} =   \frac{\langle (\Delta x_{\textrm{raw}} )^4 \rangle }{3\langle (\Delta x_{\textrm{raw}})^2 \rangle } -1 ,
\end{equation}
where $\langle .\rangle$ is the average over  the 30 min  trajectory. 
An example of this classification is shown in Fig.~\ref{fig:ngp_vs_maxdisp}, which shows both the cortical posts and the background posts for one cell, as provisionally identified at this stage. The background posts were largely clustered, but with outliers  that were mixed in with the cortical posts due to either initial mis-classification or the transient optical effects.
Cuts on both $x_{\textrm{range}}$ and the NGP, such as those shown in Fig.~\ref{fig:ngp_vs_maxdisp}, were made manually for each cell to obtain a cleaner set of background posts (those in the lower left quadrant of Fig.~\ref{fig:ngp_vs_maxdisp}) unencumbered by these effects. For some cells, comprising 15\% of the initial data set, the optical disturbances affected the majority of the background posts, and such cells were discarded. Occasional posts identified as cortical appeared in the lower left 
quadrant of  Fig.~\ref{fig:ngp_vs_maxdisp} as well. These were discarded from the data set as misidentified.

The mean trajectory of this improved set of background posts was  used to re-dedrift the trajectories of all the posts and to refine the cell-associated posts' undeflected locations. 
For the latter procedure, each line of posts running along the $[10]$, $[01]$, and $[\bar{1}1]$ directions of the hexagonal lattice was fit by an orthogonal regression on the background posts along that line. The undeflected location of each cell post was estimated as the weighted average of the intersections of the three lattice lines through the post's position. 

The procedure to bifurcate the cell-attached posts into cortical and stress-fiber-associated was repeated  using the improved resting locations to refine this classification. 
In a few cases this bifurcation procedure did not identify a clear cortical region, and these cells were discarded. 

In cases where the transient optical disturbances traversed the cell, the disturbances were masked in the cell-attached posts due to the higher activity of those posts, and so the trajectory of these disturbances as tracked in the background posts was extrapolated across the cell, and the affected cortical posts were manually discarded from the data set.

\section{Data Analysis}

{\it Rescaling of micropost trajectories by the geometric mean:}
For each post, the geometric mean, the arithmetic mean in log transform space,  at a lag time of 10~s,  $\textrm{GM}(x) = \exp(\langle \ln x( \tau = 10~\textrm{s}) \rangle)$, of the displacements was calculated. Post-to-post heterogeneity was removed by scaling each post's steps by the ratio of the mean of the geometric means of the entire ensemble of posts and the post's geometric mean.  See Eq.~1 in the main text.

{\it Tests for time-dependence of post-to-post heterogeneity and for cell-to-cell variations:} 

To quantify the post-to-post heterogeneity, the 1799 displacements (measured at $\tau$ = 1~s) from a selected post are compared to a sample of the same size drawn from the other posts of the same cell via a Kolmogorov--Smirnov (K--S) test. 

To test for potential time-dependence in the cortical fluctuation distributions due to cellular activity, for each cell type and/or substrate stiffness the van Hove distributions pooled over all the posts for each of the individual cells were computed for each 5 minute interval of the 30-minute video. A sample of 1799 steps from the distribution for one of the intervals was compared to another sample of the same size drawn from the distribution in the same cell from all other intervals using a K--S test as before.

Additionally, to test for cell-to-cell variability, we compared the van Hove distribution for each cell to the pooled distribution of the other cells of the same cell type and substrate stiffness, also via a K--S test. 

These tests showed that there was no significant time-dependence or cell-to-cell heterogeneity in our measurements.

{\it Removal of low-signal to noise posts:} In a white light imaging study, such as ours, there is the possibility that there are spurious contributions to the posts' apparent fluctuating motion due to internal cellular motions that lead to time-dependent variations in the local optical density of the cell over each micropost. To assess the potential impact of such effects, we used an approach we developed previously \cite{shi2019dissecting} to identify posts under a cell that are transiently decoupled from the cell. The trajectories of all the posts were divided into 300~s segments, and segments from posts coupled to the cell whose displacement from the resting location was low (both $\Delta x, \Delta  y < 15\,\mathrm{nm}$) for the entire duration of the segment were identified as time intervals when the post was detached from the cell.  To avoid possible effects near the edge of the cell arising from transient fluctuations in the cell's perimeter, we further focused on posts in the interior of the cell, i.e., with all neighbors also under the cell as classified by visual inspection.  The MSDs of such `detached' segments were compared to the MSDs of segments from background posts and from cortical posts (Fig.~\ref{fig:msds_detached}).  As there appeared to be some overlap in the distribution of MSDs of the detached segments with the lower part of the distribution of MSDs for cortically-attached segments,  to avoid amplifying this noise when scaling such posts' trajectories to the average geometric mean of the ensemble, only posts with a geometric mean in the uppermost quartile of the geometric mean distribution were retained (Fig.~\ref{fig:gm_distro}).

{\it Determination of displacement events' duration and direction:} To identify the subset of the largest steps in the van~Hove distributions that were associated with abrupt non-cortical events, the $x$-components of the post's motion in the thirty second interval centered around the time $t_s$ of the center of each of the largest steps measured at lag time $\tau = 10$ s was fit to a sigmoid curve of the form
\begin{equation} \label{eq:sigmoid1}
	x(t) = x_{-\infty} + \dfrac{b}{1 + e^{-(t - t_0)/\delta t}}.
\end{equation}
Steps where the widths of the fit sigmoid $\delta t$ were less than 1 second were identified as  abrupt events. Steps where the center of the fit sigmoid $t_0$ was greater than 8 s away from $t_s$ were excluded from this analysis, as such fits did not adequately characterize the step in question. 

The direction $\theta$ of the event relative to the post's undeflected location was determined from $\cos \theta = \Delta{\bf \hat{r}} \cdot {\bf \hat{r}_0}$, where $\Delta{\bf \hat{r}}$ is the unit vector in the direction of the step and ${\bf \hat{r}_0}$ is the unit vector pointing to the post's undeflected location from the position of the post at the start of the step, $t=t_s - \tau/2$. Note that with this, $\theta = 0$   for steps directly toward the resting location (See Fig.~\ref{fig:direction_separate}(a)--(b)).

{\it IEF analysis:} Here we seek to determine if a subset of large displacements in a long random trajectory are statistically consistent with the other fluctuations making up the entire trajectory.  For example, we want to detect a case where a given trajectory consists of a single stationary random process with a few large steps superimposed upon it at random times due to a distinct process. 

To begin, we construct an `average cytoquake' from the largest displacements in the ensemble of micropost trajectories.  Specifically, the 30 min trajectories of each micropost were broken into segments of length $\tau$, and a set of segments with the largest magnitude steps $|\Delta x_i| = |x(t_i+\tau/2) - x(t_i- \tau/2 )|$ within the ensemble of trajectories were identified.  For each segment, after taking $t \rightarrow t-t_i$ to center the segment on $t=0$, the portion of that post's full trajectory in the interval $-3\tau/2 \le t \le 3\tau/2$ was rescaled as
\begin{equation} \label{scaledtraj}
	x_{i,s}(t) = \frac{x(t) - x(-\frac{\tau}{2})}{\Delta x_i} - \frac{1}{2}.
\end{equation}
Note that this scaling forces each trajectory to pass through the points $(-\tau/2,-1/2)$ and $(+\tau/2,+1/2)$, no matter the sign of $\Delta x_i$. The experimental `average cytoquake' was then computed by averaging the scaled trajectories,  $\langle x_{i,s}(t)\rangle$, as shown in Fig.~3.

To construct a model for the average cytoquake or displacement trajectory above, we use the interpolation-extrapolation function (IEF) computed in Ref.(\cite{mandelbrot1968fractional}).  If $x(t)$ is a stationary random function that passes through the two points $x(\tau/2) = \Delta x/2$ and $x(-\tau/2) = -\Delta x/2$, the IEF predicts the average scaled trajectory of $x(t)$ to be

\begin{eqnarray} \label{eq:trajcorr}
	\hspace*{-0.25in}
	\textrm{IEF}(t) &=& \frac{\left\langle x(t) |  x(\frac{\tau}{2})-x(-\frac{\tau}{2})=\Delta x\right\rangle}{ \Delta x}\\
	&=& \frac{ \left\langle x(t) \left( x(\frac{\tau}{2})-x(-\frac{\tau}{2})\right) \right\rangle}{\left\langle \left( x(\frac{\tau}{2})-x(-\frac{\tau}{2}) \right)^2 \right\rangle }\\
	&=&\frac{ \left\langle \left( x(t)-x(-\frac{\tau}{2})\right)^2 \right\rangle  - \left\langle \left ( x(t)-x(\frac{\tau}{2}) \right)^2 \right\rangle }{2 \left\langle \left( x(\frac{\tau}{2})-x(-\frac{\tau}{2}) \right)^2 \right\rangle } ,
\end{eqnarray}

\noindent where $\langle x(t) |  x(\frac{\tau}{2})-x(-\frac{\tau}{2})=\Delta x\rangle$ is the conditional expectation for $x(t)$ subject to the given $\Delta x$. This equation shows that the IEF and the mean squared displacement (MSD) are mathematically related in a one to one manner. Specifically, if the MSD is a power-law function of the lag time, the predicted IEF function is given by Eq.~2 in the main text. While the above formulae were derived for a fractional Brownian process with a Gaussian van Hove distribution \cite{mandelbrot1968fractional}, we make the conventional assumption that the formulae hold for general stationary non-Gaussian random processes provided they have non-singular variance/MSD (as is the case with our data) due to the Central Limit Theorem.

{\it Estimation of error bars for IEF and MSD Exponents:} The error bars in the exponents $a_{\textrm{IEF}}$ and $a_{\textrm{MSD}}$ in Fig.~3{\em F} were determined from the
cell-to-cell variation in these quantities. In computing these error bars, the values for $a_{\textrm{MSD}}$ for each cell were weighted by the number of cortical posts in each cell. The value for $a_{\textrm{IEF}}$ for each cell was calculated using the 50 largest steps from that cell for the gradual (20 for the 3T3-H cells), and the 20 largest steps for the abrupt steps.

{\it Removal of abrupt events from displacement distributions:} For each lag time $\tau$, the direction of each displacement was classified as either inward, toward the resting location, or outward, away from it (Fig.~\ref{fig:direction_separate}{\em E}). As the abrupt events were predominantly in the `toward' category while the non-abrupt events were isotropically distributed 
we discarded all the `toward' displacements, and carried out subsequent analysis on the displacements in the `away' category.  To estimate the power carried by the abrupt events we computed the fractional energy difference 
\begin{equation}
	f_{\textrm{abrupt}} = \frac{E_{\textrm{toward}} - E_{\textrm{away}}}{E_{\textrm{toward}} + E_{\textrm{away}}},
\end{equation}
assuming that the energy carried by the fluctuations  $ E \propto \sum_i (\Delta x_i)^2$.

{\it Modeling of van Hove distributions with ETSDs:}  A maximum-likelihood estimation (MLE) technique was used to calculate the best-fit exponentially-truncated stable distribution (ETSD) $P(\Delta x)$  (Eq.~3 in the main text) 
A simplex optimization method was used to vary the parameters $(\alpha, \gamma, \lambda)$ to minimize the negative logarithm of the probability of the data being drawn from an ETSD  with those parameters. 
All calculations were done in Python, but the normalization constant $A$ in the ETSD was computed numerically in Mathematica at each iteration using the Wolfram client library for Python to ensure sufficient precision in the calculated $P(\Delta x)$ for the simplex optimization to reliably obtain the best-fit parameters.

The uncertainties in these parameters for the pooled datasets for each cell type or experimental condition were dominated by cell-to-cell variation, and so were estimated by applying this modeling procedure to each cell's van~Hove and calculating the standard error over the group of cells.

The resulting fits are shown in Figs.~4 and \ref{fig:vanhoves_allcells}.

{\it Validation of fitting function:} For all cell types and substrate stiffnesses, we fit the experimental data to the ETSD, an untruncated Stable Distribution (SD), and to a normal distribution. A KS test for performed to determine if the different forms were significantly different from one another by repeatedly (100 times) generating simulated data of each of these forms with their respective fit parameters, with as many samples as the measured data, and comparing the resulting distributions. We found that ETSD and SD forms were significantly different from the normal distribution at all but the longest lag times (i.e., with the lowest statistical power), and the SD and ETSD were significantly different at the lowest lag times in all but one of our measured conditions, namely the 3T3 cells on the substrate stiffness $k=$ 5.5 nN/$\mu$m, which had fewer cortical steps measured. 

{\it Energy estimation:}   The energy $U$ stored in a Hookean spring is given simply by $U = k \Delta x^2/2$, where $k$ is the spring constant and $\Delta x$ is the extension from an resting length.  Equivalently, the energy can be expressed in terms of the tension force $F$ via $U = F^2 / 2 k$. Two identical springs in series have the same force, and so store equal elastic energy.

When trying to understand the response of a tensed network of Hookean springs to removing a single tensed spring, we can suppose that the network can be reduced to a single equivalent spring in series having a spring constant comparable to a single spring.  Removing a single spring (e.g. by actin severing or crosslink unbinding), will then cause the remainder of the network to recoil, releasing an amount of energy comparable to that stored in the removed spring.  

When considering the motion of a mesoscopic post/spring adhered to the network, we assume that the maximum amount of work that can be done on the post, $\Delta U_{post}$, by the network rebound is that stored in the severed or unbound tensed filament.  Physically, this would correspond to a case where the removed filament was adjacent to the micropost, and the micropost had a spring constant comparable to the spring network.  The cortex and adhered microposts were shown to have somewhat lower but comparable effective spring constants in a previous study \cite{shi2019dissecting}.

The stretching stiffness of single F-actin filaments has been directly measured \cite{kojimaDirectMeasurementStiffness1994, liuMechanicsFActinCharacterized2002, matsushitaEvaluationExtensionalTorsional2010}, and found to be roughly 40 nN/$\mu$m for a 1 $\mu$m long filament and to scale inversely with filament length as expected for a slender rod.  To estimate the maximal elastic energy in a cortical actin filament we assume that they are 100 nm in length and maximally tensed to 400 pN.

\clearpage
\pagebreak

\onecolumngrid

\section{Supplementary Figures}

\renewcommand{\thefigure}{S\arabic{figure}}
\renewcommand{\theHfigure}{S\arabic{figure}}
\setcounter{figure}{0}

\begin{figure}[!h]
	\centering
	\includegraphics[width = 0.6\columnwidth]{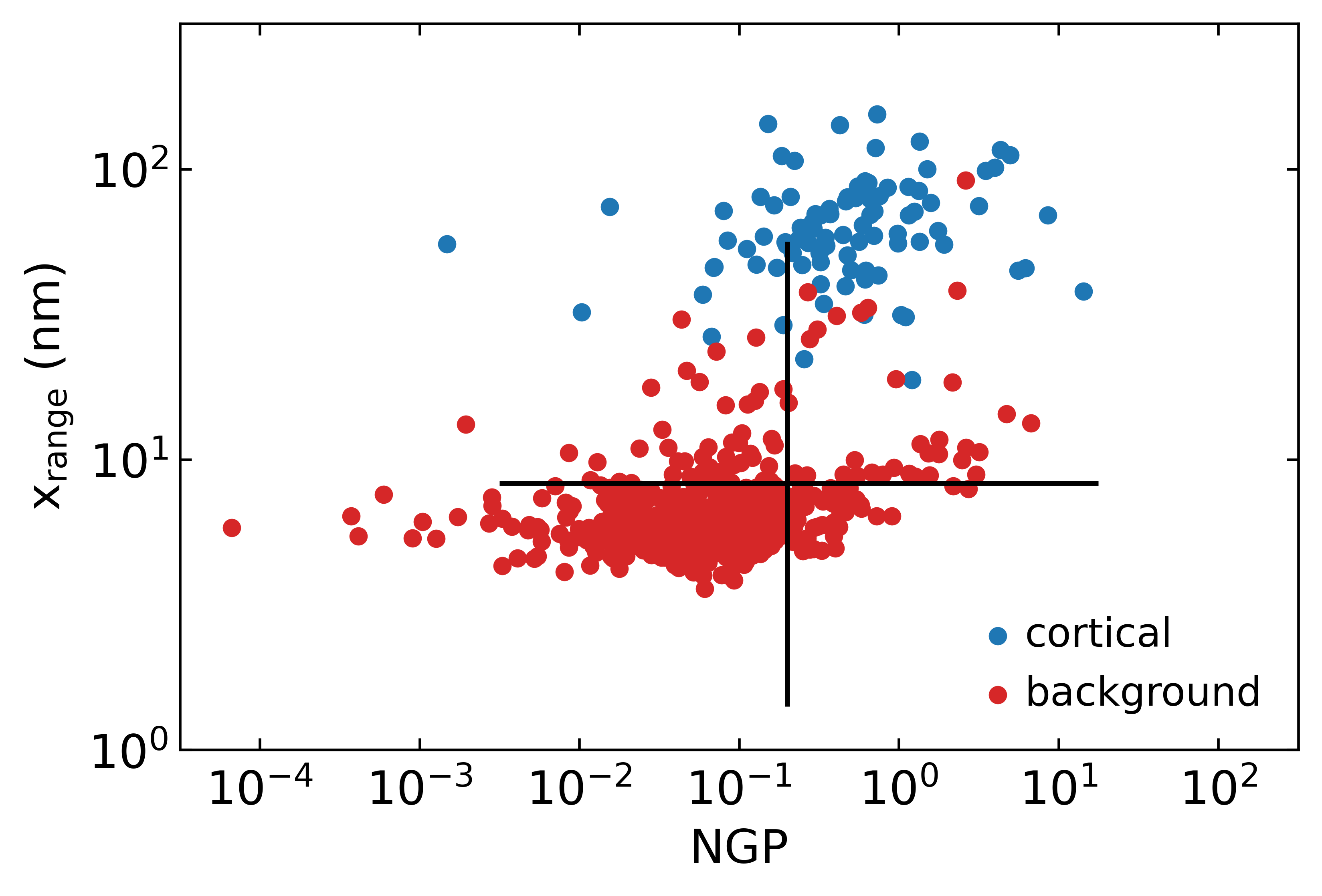}
	\caption{Scatter plot of displacement range $x_{\textrm{range}}$ vs non-Gaussian parameter (NGP) following preliminary  identification of cortical and background posts  (shown here for a single 3T3 cell on a substrate of stiffness $k = 15.8\,\textrm{nN/$\mu$m}$). Number of posts:  $N_{\textrm{cortical}} = 97$; $N_{\textrm{background}} = 660$. Based on such distributions, thresholds for $x_{\textrm{range}}$ and the NGP (solid lines)   were manually chosen for each cell to isolate a final set of background posts (lower left quadrant)  that was subsequently used to dedrift the posts' trajectories and to obtain the cortical posts' undeflected positions.}
	\label{fig:ngp_vs_maxdisp}
\end{figure}

\begin{figure}
	\centering
	\includegraphics[width = 0.8\columnwidth]{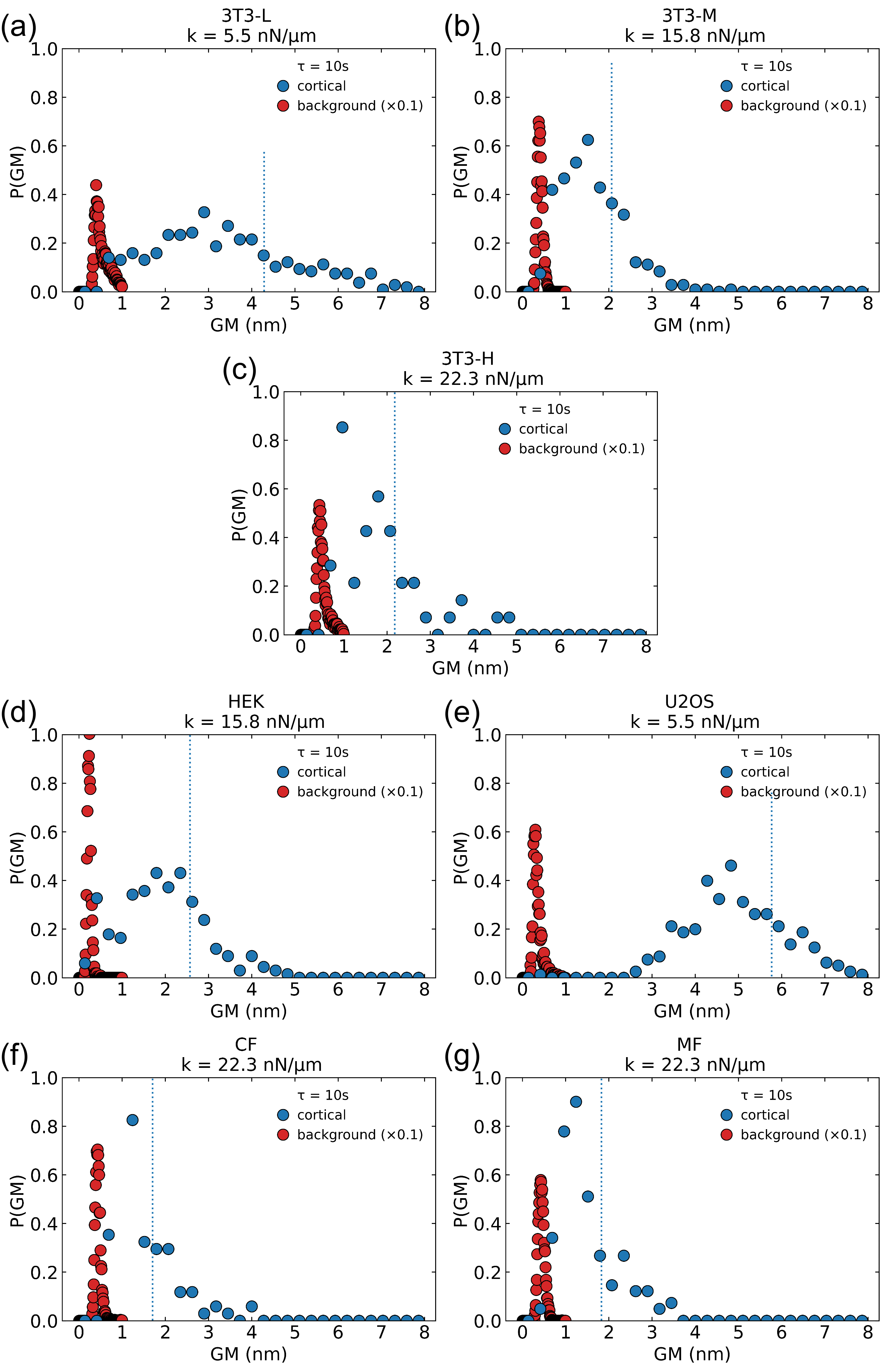}
	\caption{The probability distribution $P(GM)$ of the geometric means GM of the cortical posts compared to background posts for all cell types and substrate stiffnesses. In all cases, the distribution is much broader for the cortical posts. Only the posts with geometric mean in the uppermost quartile, to the right of the dashed line, are retained for further analysis, to avoid undue amplification of experimental noise.}
	\label{fig:gm_distro}
\end{figure}

\begin{figure}
	\centering
	\includegraphics[width = 0.55\columnwidth]{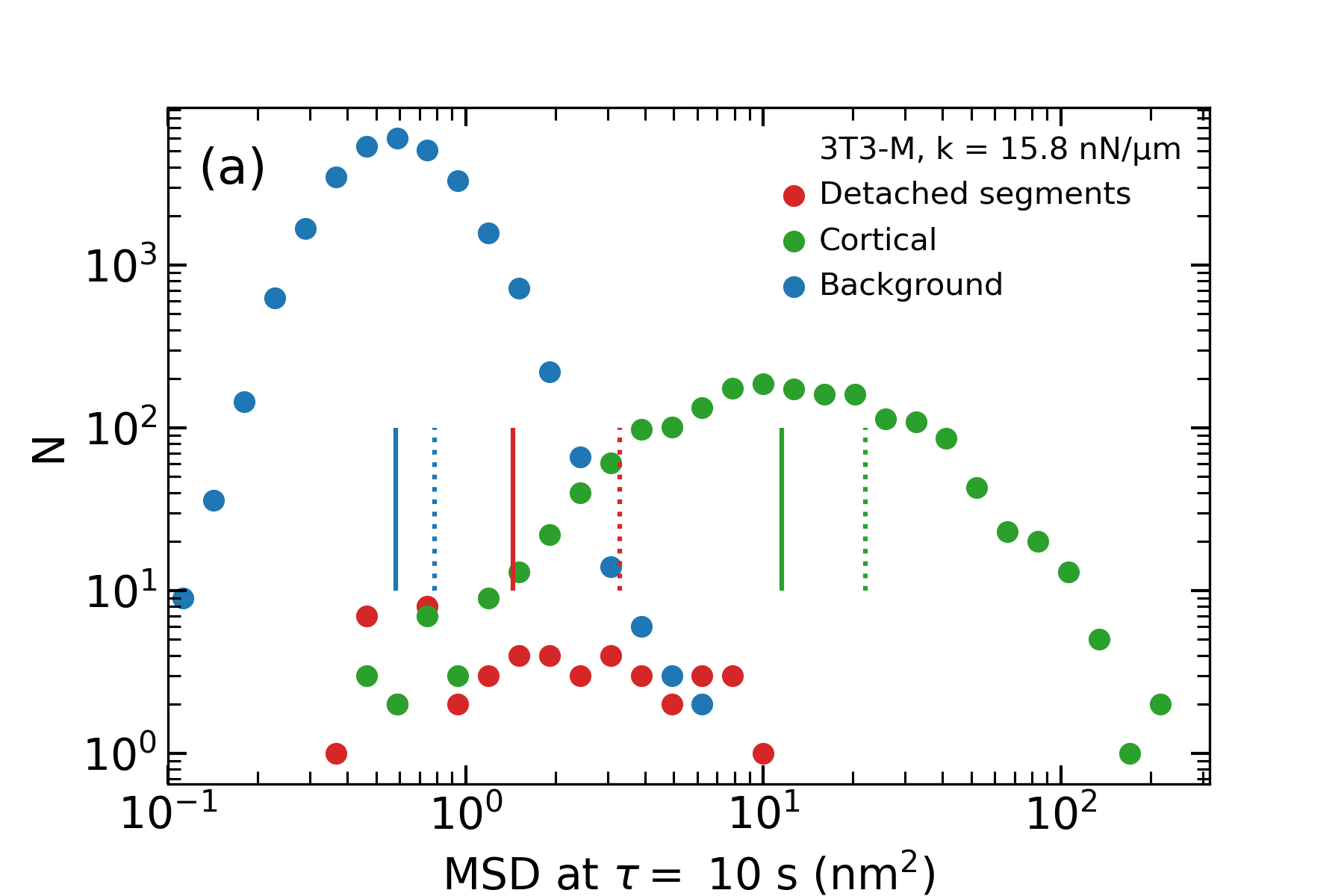}
	\includegraphics[width = 0.55\columnwidth]{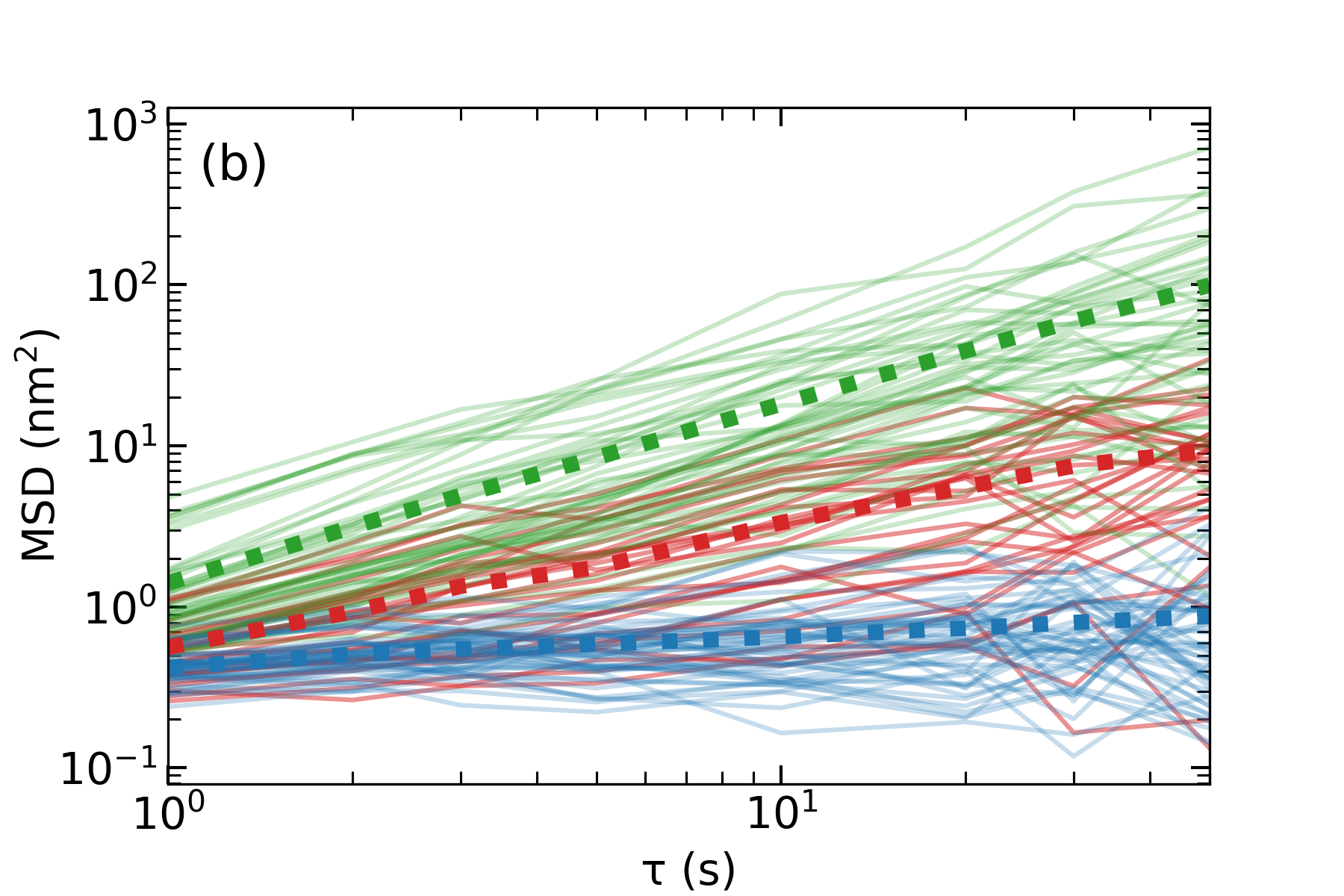} 
	\includegraphics[width = 0.4\columnwidth]{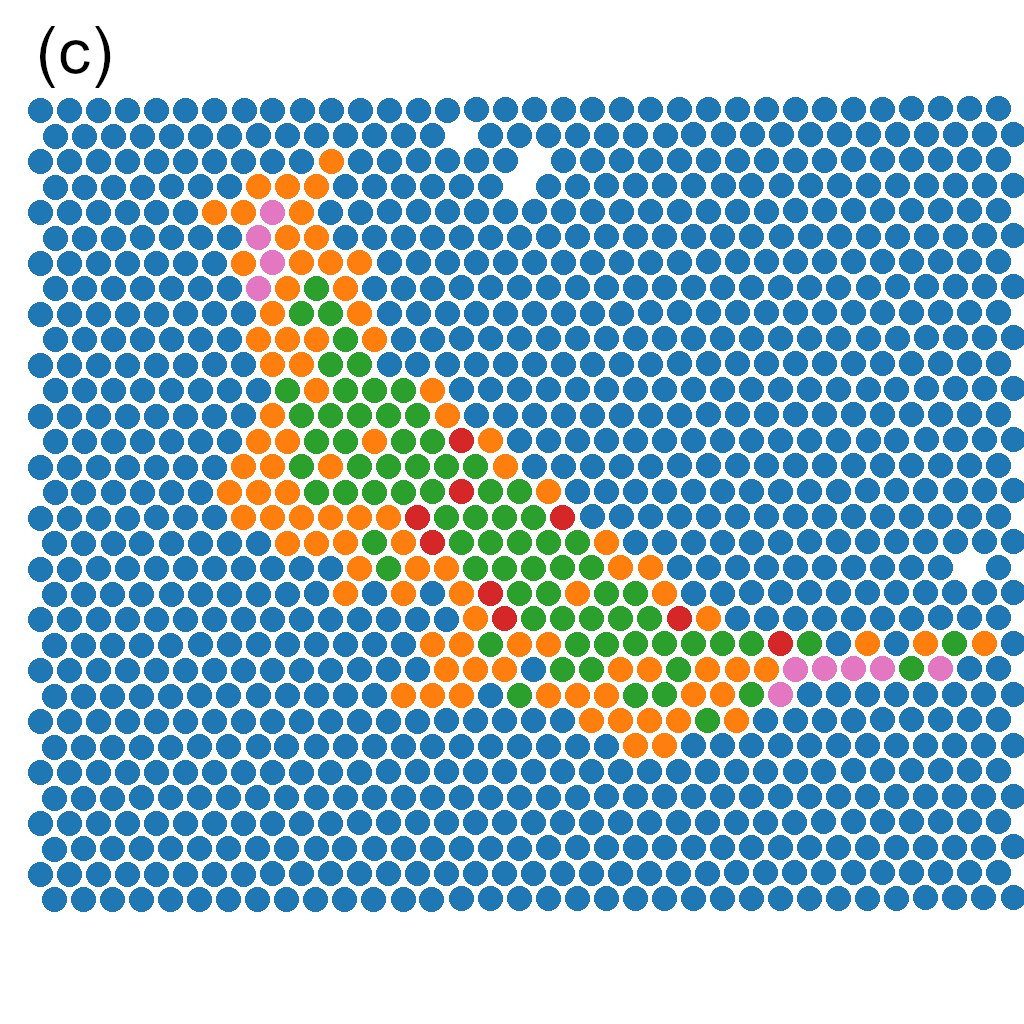}          
	
	\caption{To estimate the effect of optical noise due to transient intracellular rearrangments, we identified 300~s segments in the trajectories of posts in the interior of the cell, where the displacement was small ($\Delta x, \Delta  y < 15\,\mathrm{nm}$), and the post was likely transiently decoupled from the cell. There were 30 of these segments from our ensemble of 3T3-M cells ($k = 15.8\,\textrm{nN/$\mu$m}$). (a) Comparison of the MSDs of the transiently detached segments, computed at $\tau = 10$~s, to segments of the same length from posts coupled to the cortex (2016 segments), and from background posts (28,278 segments) shows that is an overlap between the distributions of MSDs of the cortical segments and the detached segments. The locations of the medians (solid lines) and lower bounds of the upper quartiles (dotted lines) are also shown. (b)~The MSDs of the different segments, plotted vs. $\tau$, showing the detached segments, and 50 segments each sampled from the cortical posts and background posts, show that there is an overlap between cortical segments and detached segments across a wide range of lag times. (c) An example of a cell with posts that contained detached segments marked in red. The other cell posts are colored according to their classification by traction force: cortical posts in green, stress fibers in pink, and posts intermediate or otherwise not categorized as either in orange. Background posts are shown in blue. Note that there are gaps in the array from posts that are excluded due to imperfections in the mPAD fabrication process.
	}

	\label{fig:msds_detached}
\end{figure}

\begin{figure}
	\centering
	\centerline{\includegraphics[width = 1\columnwidth]{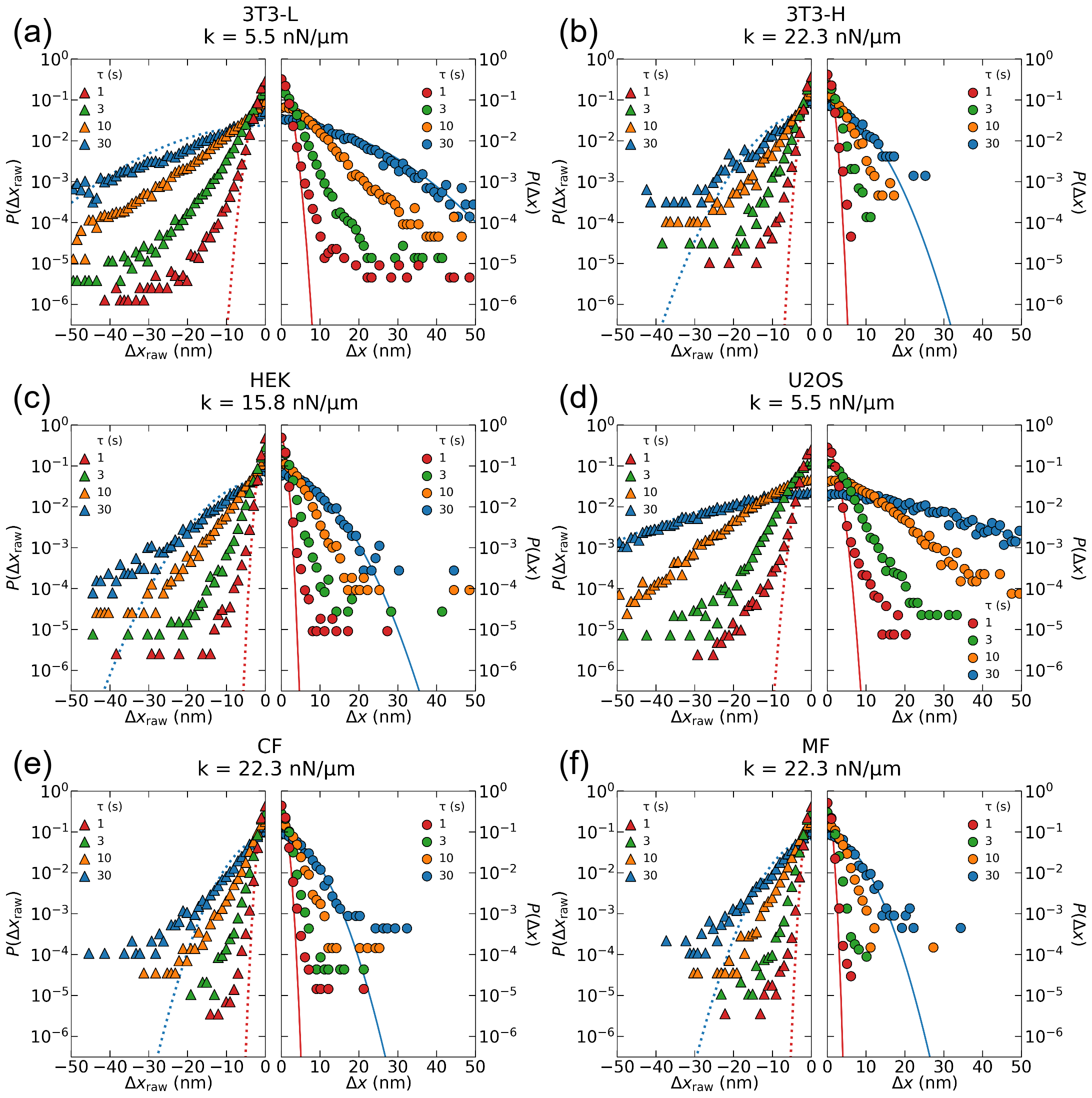}}
	\caption{The effect of the GM rescaling procedure to remove static heterogeneity (described in the main text) on the distributions of cortical displacements for various cell types, and for 3T3s on different substrate stiffnesses: (a) 3T3-L (361 posts from 11 cells; $k = 5.5\,\textrm{nN/$\mu$m}$). (b) 3T3-H (40 posts from 7 cells; $k = 22.3\,\textrm{nN/$\mu$m}$). (c) HEK cells (225 posts from 8 cells). (d) U2OS cells (266 posts from 6 cells). (e) Cardiac fibroblasts (CF) (119 posts from 9 cells). (f) Cardiac myofibroblasts (MF) (125 posts from 11 cells). In each case, the left panel is the distribution of measured displacements $P(\Delta x_{\textrm{raw}})$ for the $x$-component of the post trajectories at fixed lag time $\tau$. Only the $\Delta x_{\textrm{raw}}\le 0$ half of each symmetric distribution is shown, to facilitate comparison with the right panels, which are the Van~Hove distributions $P(\Delta x)$ following rescaling  via Eq.~1. In the right panels, only the $\Delta x \ge 0$ half of the symmetric scaled distribution is shown, and the data in the right panels only include steps from posts in the uppermost quartile of the trajectories' geometric means. The lines in each panel are best-fit Gaussians at $\tau = 1\,\textrm{s}$ and $\tau = 30\,\textrm{s}$. }
	\label{fig:rescaled_vanhove_allcells}
\end{figure}

\begin{figure}
	\centering
	\includegraphics[width = 0.7\columnwidth]{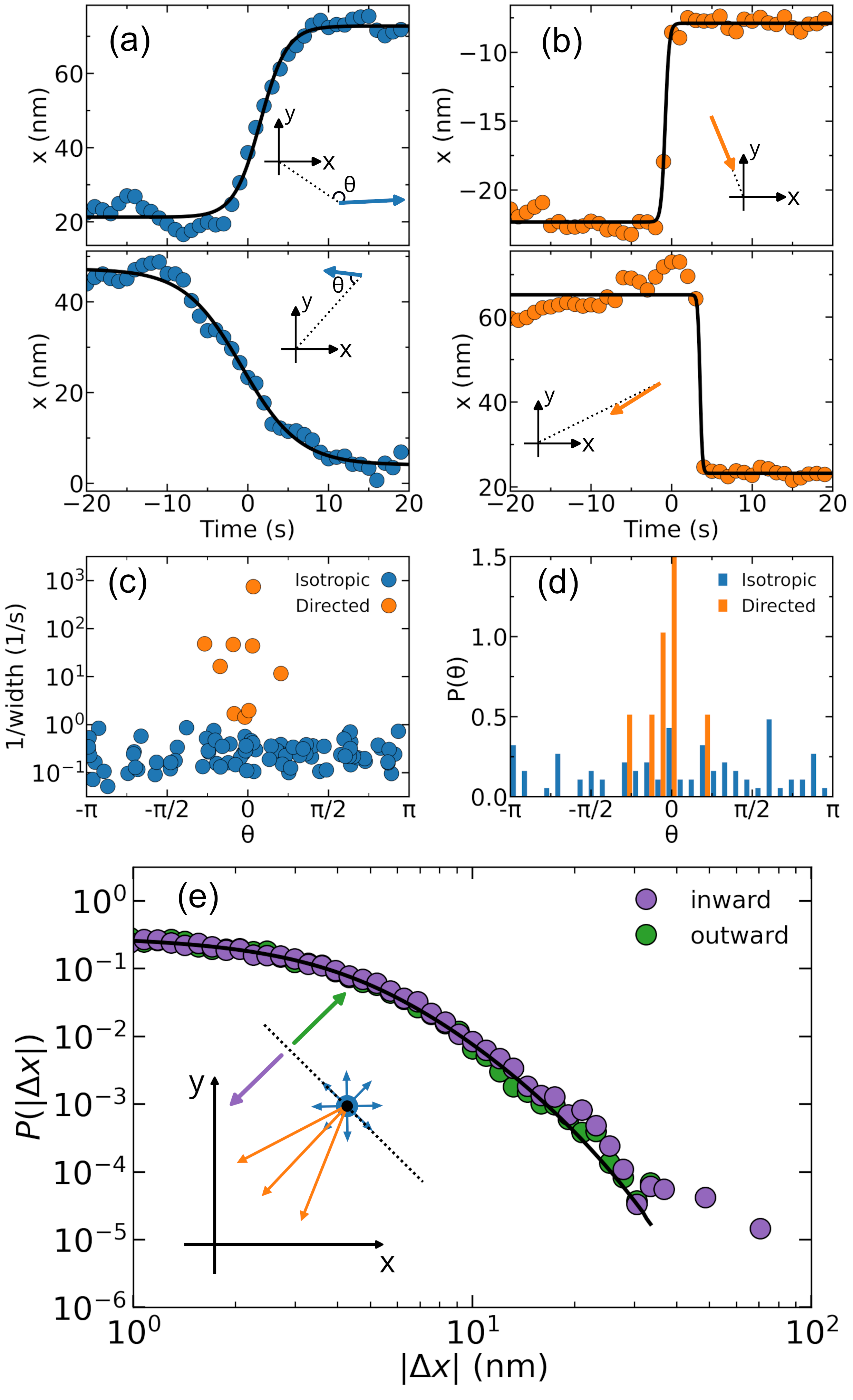}
	\caption{The dominant isotropic fluctuations in the cortex were isolated by separating steps by their direction relative to the posts' resting locations. The data shown are for 3T3-M cells ($k = 15.8\,\textrm{nN/$\mu$m}$). (a) Events extended in time with displacements that were isotropically directed in space. (b) Abrupt events directed toward the resting location of the posts. Only the $x$-components of the post trajectories are shown. The arrows in the insets show the magnitude and direction of the event in the interval $-5$ s $\le t\le$ 5 s relative to the post's undeflected position. (c) Inverse widths vs. direction and classification for the 100 largest events. (d) The angular distribution of the two kinds of events ($\theta = 0$ is toward the resting location.) (e) Van Hove distributions at $\tau = 10$ s for events separated by direction relative to the post's resting location.  To remove the abrupt, directed events, we separated the steps by angle relative to the posts' resting locations, retaining only those with $|\theta| \ge \pi/2$  This figure highlights the importance of this separation, as inclusion of the large directed events would distort the tails of the cortical fluctuation distribution. The solid line is the ETSD fit to the `outward' data (See Fig.~4 in the main text.) As indicated in Panel (d), the separation by angle is imperfect, and some residual abrupt events are perforce included in our final data set. However, as the isotropic events were found to carry 80\% of the fluctuation energy, this contribution was likely small.}
	\label{fig:direction_separate}
\end{figure}

\begin{figure}
	\centering
	\includegraphics[width = 1\columnwidth]{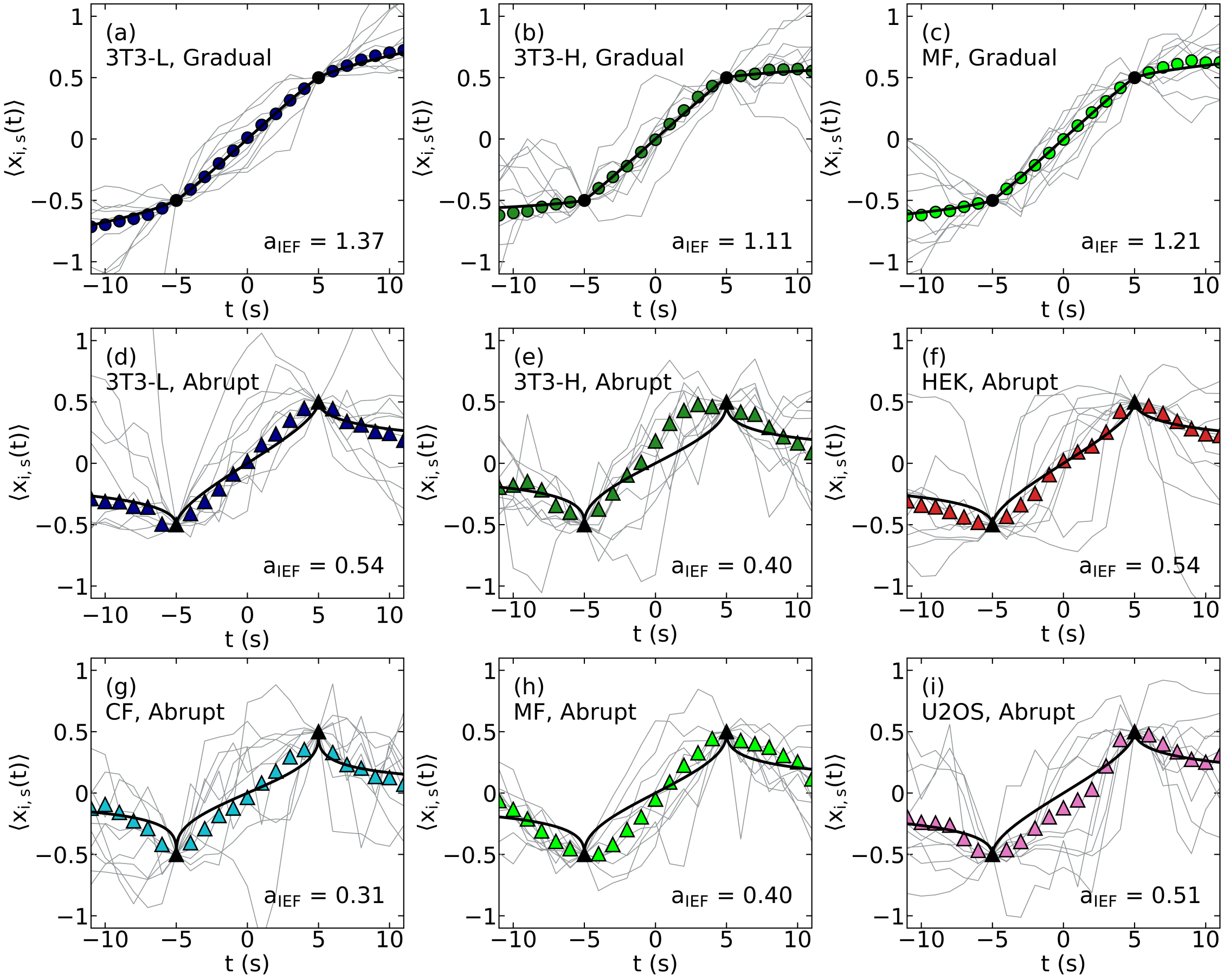}
	\caption{Averaged scaled trajectories $\langle x_{i,s}(t)\rangle$ for large displacements compared to the interpolation-extrapolation function (IEF). (a)--(c)~$\langle x_{i,s}(t)\rangle$ for the $200$ largest non-abrupt displacements at $\tau = 10$ s, for 3 different cell types/conditions, as described in the text. (d)--(i))~$\langle x_{i,s}(t)\rangle$ for the $40$ largest abrupt displacements for 6 different cell types/conditions.   Solid lines are fits to the IEF, Eq.~2 in the main text. The results for the exponent $a_{\textrm{IEF}}$ are included in Fig.~3{\em F}. Ten individual scaled trajectories $x_{i,s}(t)$ are superimposed in each case (grey). The micropost stiffnesses were: 3T3-L and U2OS: $k=$ 5.5 nN/$\mu$m;  3T3-M and HEK: $k=$ 15.8 nN/$\mu$m; 3T3-H, CF and MF: $k=$ 22.3 nN/$\mu$m. The 3T3-H cells had fewer cortical steps  measured, and only the 40 largest trajectories were averaged for calculating  $a_{\textrm{IEF}}$.}  
	\label{fig:ief_si}
\end{figure}

\begin{figure}
	\centering
	\includegraphics[width = 0.6\columnwidth]{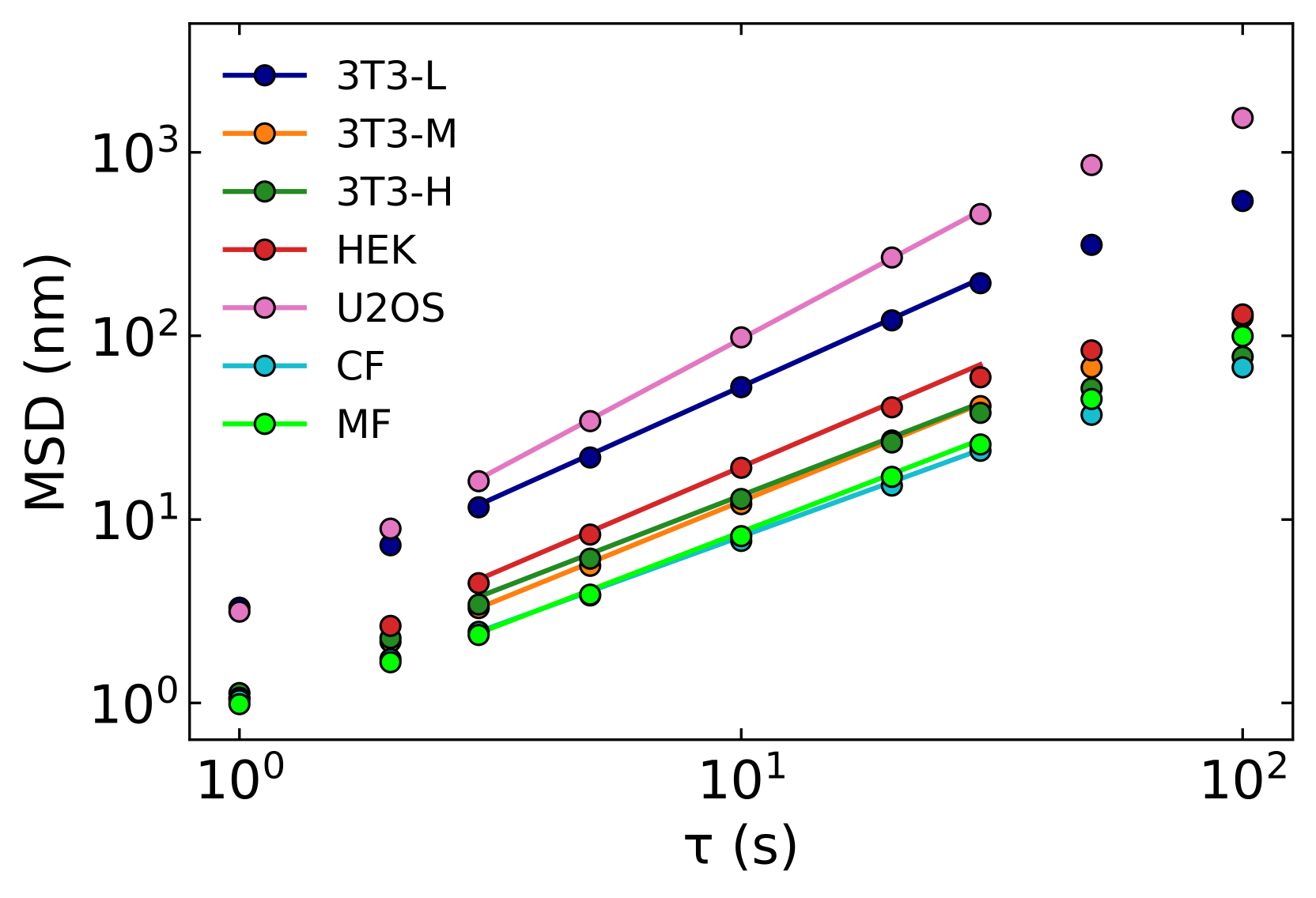}
	\caption{The MSDs of the displacements for different cell types and substrate stiffnesses were determined using the variances of the van Hove distributions.
		The MSD exponents $a_{\textrm{MSD}}$ were determined from fits to  $\ln \textrm{MSD} = a_{\textrm{MSD}} \ln \tau + b$ over the range 3\,s~$\le \tau \le $~30\,s. The resulting values for $a_{\textrm{MSD}}$ are shown in Fig.~3{\em F}.}
	\label{fig:variance_slope}
\end{figure}

\begin{figure}
	\centering
	\includegraphics[width = 0.75\columnwidth]{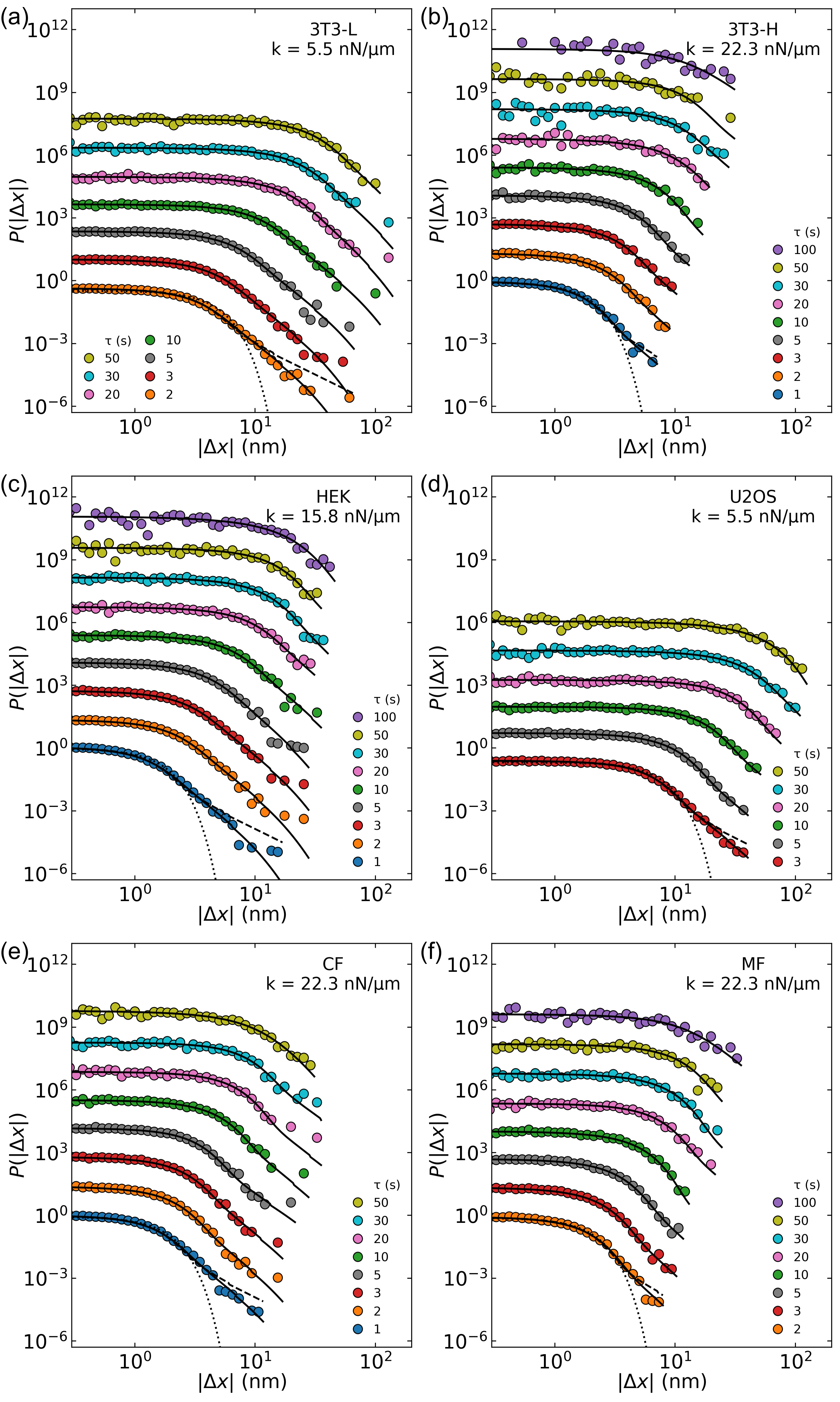}        
	\caption{The van~Hove distributions of the cortical posts at different lag times for different cell types and substrate stiffnesses are all well-described by our model of ETSD plus noise (solid lines). The resulting ETSD parameters $\alpha$, $\gamma$ and $\lambda$ are shown in Fig.~5 in the main text. The best-fit stable distributions (dashed lines) shown at $\tau = 1$ s do not capture the behavior near the tails in most of our measured conditions. Best-fit Gaussians are shown at $\tau = 1$ s for reference (dotted lines). The distributions for $\tau > 1$ s are progressively offset by factors of $10^{1.5}$ in each panel for clarity. }
	\label{fig:vanhoves_allcells}
\end{figure}

\end{document}